\documentclass[sigconf,screen]{acmart}
\setcopyright{rightsretained}
\acmPrice{}
\acmDOI{10.1145/3533767.3534220}
\acmYear{2022}
\copyrightyear{2022}
\acmSubmissionID{issta22main-p174-p}
\acmISBN{978-1-4503-9379-9/22/07}

\acmConference[ISSTA '22]{Proceedings of the 31st ACM SIGSOFT International Symposium on Software Testing and Analysis}{July 18--22, 2022}{Virtual, South Korea}
\acmBooktitle{Proceedings of the 31st ACM SIGSOFT International Symposium on Software Testing and Analysis (ISSTA '22), July 18--22, 2022, Virtual, South Korea}

\usepackage{microtype}
\usepackage{xspace}
\usepackage{adjustbox}
\usepackage{amsmath,amsfonts}
\usepackage{algorithm}
\usepackage[noend]{algpseudocode}
\usepackage{graphicx}
\usepackage{textcomp}
\usepackage{xcolor}
\usepackage{booktabs}
\usepackage{listings}
\usepackage{csquotes}
\usepackage{multirow}
\usepackage{siunitx}
\usepackage{enumitem}

\usepackage{makecell}

\usepackage{balance}
\usepackage{tabularx}
\usepackage{comment}
\usepackage{tikz}

\usepackage[caption=false]{subfig}

\def\1{\bm{1}}

\DeclareMathAlphabet{\mathsfit}{\encodingdefault}{\sfdefault}{m}{sl}
\SetMathAlphabet{\mathsfit}{bold}{\encodingdefault}{\sfdefault}{bx}{n}

\newcommand{\E}{\mathbb{E}}

\newcommand{\argmin}[1]{\underset{#1}{\operatorname{arg}\,\operatorname{min}}\;}

\newcommand{\tool}{\rev{DocTer}\xspace}

\newcommand{\q}[1]{``{#1}''}

\newcommand{\rev}[1]{{#1}}

\newcommand{\dtype}{\textit{dtype}\xspace}
\newcommand{\dtypes}{\textit{dtypes}\xspace}
\newcommand{\shape}{\textit{shape}\xspace}
\newcommand{\datastructure}{\textit{structure}\xspace}
\newcommand{\validvalue}{\textit{valid value}\xspace}

\newcommand{\code}[1]{\texttt{\small#1}\xspace}

\newcommand{\todoc}[2]{{\textcolor{#1}{\textbf{#2}}}}

\newcommand{\todored}[1]{{\todoc{red}{\textbf{[[#1]]}}}}

\newcommand{\todo}[1]{\todored{TODO: #1}}

\definecolor{light-gray}{gray}{0.7}

\renewcommand{\todoc}[2]{\relax}

\newcommand{\tftotalconstraint}{5,908\xspace}
\newcommand{\pytorchtotalconstraint}{3,201\xspace}
\newcommand{\mxnettotalconstraint}{6,926\xspace}
\newcommand{\totalconstraint}{16,035\xspace}

\newcommand{\tfnumpat}{665\xspace}
\newcommand{\pytorchnumpat}{398\xspace}
\newcommand{\mxnetnumpat}{275\xspace}
\newcommand{\totalnumpat}{1,338\xspace}

\newcommand{\tfapiwithconstraint}{911\xspace}
\newcommand{\pytorchapiwithconstraint}{498\xspace}
\newcommand{\mxnetapiwithconstraint}{1,006\xspace}
\newcommand{\totalapiwithconstraint}{2,415\xspace}

\newcommand{\tfnumdoc}{190\xspace}  
\newcommand{\tfnumconstraint}{350\xspace}  
\newcommand{\pytorchnumdoc}{93\xspace}
\newcommand{\pytorchnumconstraint}{170\xspace}
\newcommand{\mxnetnumdoc}{320\xspace}
\newcommand{\mxnetnumconstraint}{363\xspace}
\newcommand{\tolnumdoc}{603\xspace}
\newcommand{\tolnumconstraint}{883\xspace}

\newcommand{\tfpredtype}{93.0}
\newcommand{\tfpreds}{78.9}
\newcommand{\tfpreshape}{89.1}
\newcommand{\tfprevalidvalue}{87.5}
\newcommand{\tfpretotal}{90.0}

\newcommand{\tfrecalldtype}{82.3}
\newcommand{\tfrecallds}{88.2}
\newcommand{\tfrecallshape}{74.5}
\newcommand{\tfrecallvalidvalue}{47.7}
\newcommand{\tfrecalltotal}{74.8}

\newcommand{\tffscoredtype}{87.3}
\newcommand{\tffscoreds}{83.3}
\newcommand{\tffscoreshape}{81.2}
\newcommand{\tffscorevalidvalue}{61.8}
\newcommand{\tffscoretotal}{81.7}

\newcommand{\ptpredtype}{78.1}
\newcommand{\ptpreds}{85.7}
\newcommand{\ptpreshape}{80.0}
\newcommand{\ptprevalidvalue}{66.7}
\newcommand{\ptpretotal}{78.4}

\newcommand{\ptrecalldtype}{79.4}
\newcommand{\ptrecallds}{90.0}
\newcommand{\ptrecallshape}{76.9}
\newcommand{\ptrecallvalidvalue}{60.0}
\newcommand{\ptrecalltotal}{77.4}

\newcommand{\ptfscoredtype}{78.7}
\newcommand{\ptfscoreds}{87.8}
\newcommand{\ptfscoreshape}{78.4}
\newcommand{\ptfscorevalidvalue}{63.2}
\newcommand{\ptfscoretotal}{77.9}

\newcommand{\mxpredtype}{92.9}
\newcommand{\mxpreds}{91.7}
\newcommand{\mxpreshape}{76.1}
\newcommand{\mxprevalidvalue}{90.0}
\newcommand{\mxpretotal}{87.9}

\newcommand{\mxrecalldtype}{81.9}
\newcommand{\mxrecallds}{90.2}
\newcommand{\mxrecallshape}{79.8}
\newcommand{\mxrecallvalidvalue}{60.0}
\newcommand{\mxrecalltotal}{82.4}

\newcommand{\mxfscoredtype}{87.0}
\newcommand{\mxfscoreds}{92.4}
\newcommand{\mxfscoreshape}{77.9}
\newcommand{\mxfscorevalidvalue}{72.0}
\newcommand{\mxfscoretotal}{85.1}

\newcommand{\pretotal}{85.4}
\newcommand{\predtype}{88.0}
\newcommand{\preds}{85.4}
\newcommand{\preshape}{81.7}
\newcommand{\prevalidvalue}{81.4}

\newcommand{\recalltotal}{78.2}
\newcommand{\recalldtype}{81.2}
\newcommand{\recallds}{89.5}
\newcommand{\recallshape}{77.1}
\newcommand{\recallvalidvalue}{55.9}

\newcommand{\fscoretotal}{81.6}
\newcommand{\fscoredtype}{84.3}
\newcommand{\fscoreds}{87.8}
\newcommand{\fscoreshape}{79.2}
\newcommand{\fscorevalidvalue}{65.7}

\newcommand{\prevereval}{81.9}
\newcommand{\recallvereval}{77.7}
\newcommand{\fscorevereval}{79.7}

\newcommand{\manhour}{36\xspace}
\newcommand{\baseNewVeri}{34\xspace}
\newcommand{\baseNewAll}{41\xspace}
\newcommand{\baseAll}{59\xspace}

\newcommand{\oursNewVeri}{54\xspace}
\newcommand{\oursNewAll}{63\xspace}
\newcommand{\oursAll}{94\xspace}
\newcommand{\oursCIAll}{75\xspace}
\newcommand{\oursVIAll}{81\xspace}

\newcommand{\oursMinusbase}{52\xspace}
\newcommand{\missingBugBase}{7\xspace}
\newcommand{\knownBugFixed}{31\xspace}

\newcommand{\numAPICrashTF}{114\xspace}
\newcommand{\numAPICrashPT}{28\xspace}
\newcommand{\numAPICrashMX}{32\xspace}
\newcommand{\numAPICrashTotal}{174\xspace}
\newcommand{\numAPICrashBase}{108\xspace}
\newcommand{\numAPICrashDep}{12\xspace}

\newcommand{\newBugFixed}{49\xspace}
\newcommand{\newBugFixedC}{19\xspace}
\newcommand{\newBugFixedPy}{11\xspace}
\newcommand{\newBugFixedboth}{7\xspace}
\newcommand{\newBugFixedPytotal}{26\xspace}
\newcommand{\newBugFixedCtotal}{18\xspace}
\newcommand{\newBugFixedunknown}{12\xspace}
\newcommand{\newBugConfirmed}{5\xspace}

\newcommand{\docBugAll}{43\xspace}
\newcommand{\docBugVeri}{39\xspace}
\newcommand{\docBugFixed}{35\xspace}
\newcommand{\docBugConfirmed}{4\xspace}
\newcommand{\docBugFormat}{11\xspace}
\newcommand{\docBugInconsis}{29\xspace}
\newcommand{\docBugDep}{3\xspace}

\newcommand{\advcite}{\cite{tensorfuzz, dlfuzz, deep-hunter, grammar-based-testing, wang2018simply, deepmutationpp, 2020FuzzTB}}

\newcommand\encircle[1]{%
  \tikz[baseline=(X.base)] 
    \node (X) [draw, shape=circle, inner sep=0, fill=black, text=white] {\strut #1};%
}

\algblockdefx[Foreach]{Foreach}{EndForeach}[1]{\textbf{foreach} #1 \textbf{do}}{\textbf{end foreach}}
\pdfoutput=1
\begin{document}

\title{
\tool: Documentation-Guided Fuzzing for Testing Deep Learning API Functions
}

\author{Danning Xie}
\affiliation{%
  \institution{Purdue University}
  \city{West Lafayette}
  \state{IN}
  \country{USA}
}
\email{xie342@purdue.edu}

\author{Yitong Li}
\affiliation{%
  \institution{University of Waterloo}
  \city{Waterloo}
  \state{ON}
  \country{Canada}
}
\email{ytnli95@gmail.com}

\author{Mijung Kim}
\authornote{The work was completed when Mijung Kim was at Purdue University.}
\affiliation{%
  \institution{Ulsan National Institute of Science and Technology}
 \city{Ulsan}
 \state{}
 \country{South Korea}
}
\email{mijungk@unist.ac.kr}

\author{Hung Viet Pham}
\affiliation{%
  \institution{University of Waterloo}
  \city{Waterloo}
  \state{ON}
  \country{Canada}
}

\email{hvpham@uwaterloo.ca}

\author{Lin Tan}
\authornote{Corresponding author.}
\affiliation{%
  \institution{Purdue University}
  \city{West Lafayette}
  \state{IN}
  \country{USA}
}
\email{lintan@purdue.edu}

\author{Xiangyu Zhang}
\affiliation{%
  \institution{Purdue University}
  \city{West Lafayette}
  \state{IN}
  \country{USA}
}
\email{xyzhang@cs.purdue.edu}

\author{Michael W. Godfrey}
\affiliation{%
  \institution{University of Waterloo}
  \city{Waterloo}
  \state{ON}
  \country{Canada}
}
\email{migod@uwaterloo.ca}

\begin{abstract}

Input constraints are useful for many software development tasks. 
For example,  input constraints of a function enable the generation of valid inputs, i.e., inputs that follow these constraints, to test  the function deeper. 
API functions of deep learning (DL) libraries have DL-specific input constraints, which are  described informally in the free-form API documentation. Existing constraint-extraction techniques are ineffective for extracting DL-specific input constraints.

To fill this gap, we design and implement a new technique---\emph{\tool}---to analyze API documentation to extract DL-specific input constraints for DL API functions. 
\tool features a novel algorithm that  automatically constructs rules to  extract API parameter constraints from
syntactic patterns in the form of dependency parse trees  
of API descriptions. These rules are then applied to a large volume of API documents in popular DL libraries to extract their input parameter constraints.
To demonstrate the effectiveness of the extracted constraints, \tool{}  uses the constraints to 
enable the automatic generation of valid  and invalid inputs to test DL API functions.

Our evaluation on three popular DL libraries (TensorFlow, PyTorch, and MXNet) shows that \tool's precision in 
extracting input constraints is \pretotal\%.
\tool 
detects \oursAll bugs from \numAPICrashTotal API functions, including one previously unknown \textbf{security vulnerability} that is now documented in the CVE database, while a baseline technique without input constraints detects only \baseAll bugs. 
Most (\oursNewAll) of  the \oursAll bugs are previously unknown, 
\oursNewVeri of which have  been fixed or confirmed  by developers after we report them. 
In addition, \tool detects \docBugAll inconsistencies in documents, \docBugVeri of which are fixed or confirmed.

\end{abstract}

\ccsdesc[500]{Software and its engineering~Software reliability}
\ccsdesc[500]{Software and its engineering~Software testing and debugging}

\keywords{
text analytics, 
testing, 
test generation, 
deep learning}

\maketitle

\section{Introduction}
\label{sec:intro}

Input constraints are useful for various software development tasks~\cite{wong2015dase, c2s, jdoctor, grammarFuzzing, CESE}.
For example, input constraints of a function enable the generation of valid inputs, i.e., inputs that follow these constraints, to test the function deeper.
API functions of DL libraries expect their input arguments to follow constraints, many of which are DL-specific. For example, 
one parameter 
\code{input} of the PyTorch API function \code{torch.as\_strided} has to be a tensor.
A \emph{tensor} is represented using an n-dimensional array, where $n$ is a non-negative integer. Any input that cannot be interpreted as a tensor (e.g., a Python list) is invalid.
Many such DL-specific input constraints are described informally in free-form API documentation. 
The availability of such DL API documentation presents a great opportunity to  automatically extract DL-specific constraints for better testing and other software development tasks.

Specifically, DL libraries' API functions require two types of constraints for their input arguments: (1)  data structures and  (2)  properties of these data structures.
First, DL libraries often require their input arguments to be specific \emph{data structures} such as lists, tuples, and tensors 
to perform numerical computations.
For example, \code{input} of the PyTorch API function \code{torch.as\_strided} has to be a tensor as dictated by its API document.
Any input that cannot be interpreted as a tensor (e.g., a Python list)
is rejected by the function's input validity check.
Such invalid inputs exercise only the input validity checking code, failing to test the core functionality of the API function. 
To test \code{as\_strided}'s core functionality, 
a testing technique needs to generate a tensor object for the \code{input} parameter.

\begin{figure}[t]
    
    \centering
    \subfloat[API Document\label{fig:API_example}]{%
       \includegraphics[clip=true,trim=0mm 1.5mm 0mm 0mm,width=\linewidth]{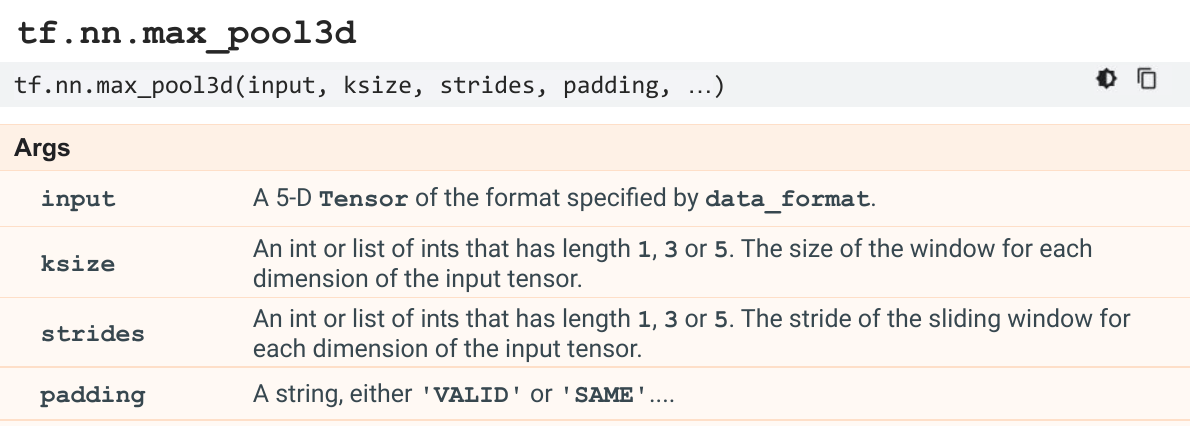}}
       \vspace{-3mm}
       \\

    \subfloat[Extracted constraints\label{fig:API_example_constraint}]{%
       \includegraphics[clip=true,trim=0mm 0.5mm 0mm 0mm,width=0.56\linewidth]{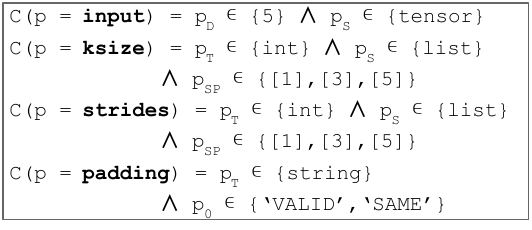}}
    \subfloat[Bug-triggering input\label{fig:API_example_input}]{%
       \includegraphics[clip=true,trim=0mm 0.5mm 0mm 0mm,width=0.42\linewidth]{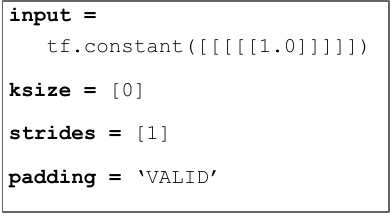}}
       \vspace{-3mm}
       \\
    \subfloat[Bug fix in \footnotesize{\texttt{pooling\_ops\_3d.cc}}\label{fig:API_example_fix}]{%
       \includegraphics[clip=true,trim=0mm 4mm 0mm 0mm,width=\linewidth]{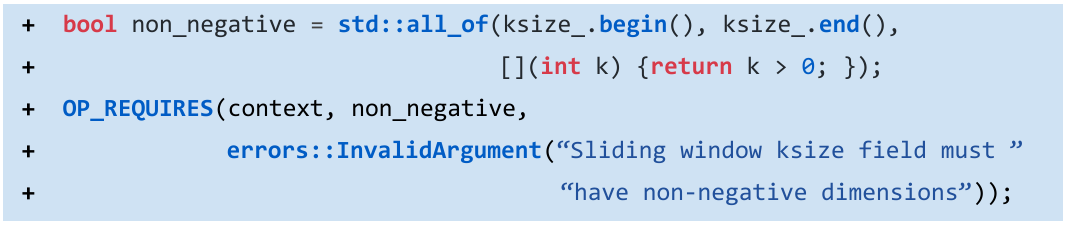}}
       
    \vspace{-3mm}
	\caption{TensorFlow document   
 helps our tool detect a bug that was fixed after we reported it to TensorFlow 
 developers.
	}
	\label{fig:API_example_all}
	\vspace{-5mm}
\end{figure}

Second, API functions of DL libraries require their arguments to satisfy  specific \emph{properties} of  data structures. 
Generating a correct data structure with incorrect properties often fails the input validity checking of the DL API functions.
They often require
two common properties of a data structure---\dtype and \shape.
Property \dtype specifies the data type of the data structure (e.g., \code{int32}, \code{float64}, and \code{String}).
In Fig.~\ref{fig:API_example}, 
the \dtype of the parameter \code{padding}  should be \code{String}.
Property \shape specifies the length of each dimension of the data structure.
For example, a \shape of $3\times4$ matrix is a 2-dimensional tensor with the first dimension of 3 elements and the second dimension of 4 elements.  
As another example, Fig.~\ref{fig:API_example} shows the document for TensorFlow API \code{tf.nn.max\_pool3d}, which indicates that the parameter \code{input} should be a tensor of 5 dimensions, with the size of each dimension unspecified.
Similarly, any inputs that violate these \dtype or \shape requirements are rejected, failing to test the core functionality of the API function.

\emph{While existing techniques can extract  constraints from code or software text (e.g., comments and  documents), they are insufficient for extracting DL-specific constraints. }
Specifically, while Pytype~\cite{pytype} infers data types from Python code, it cannot precisely infer types for DL libraries because it cannot analyze across Python and C++ code. In addition, it cannot extract numerical constraints such as \shape and \textit{range}.
Existing techniques that derive constraints from software text 
extract different types of constraints that are not DL-specific, such as 
exceptions~\cite{jdoctor, c2s}, 
command-line options and file formats~\cite{wong2015dase},
locking~\cite{icomment}, call-relations~\cite{icomment, RTFM-advance}, interrupts~\cite{acomment}, nullness~\cite{tcomment, zhou2017analyzing}
and inheritance relations~\cite{zhou2017analyzing}.
Although some~\cite{jdoctor, zhou2017analyzing, c2s, RTFM-advance} can extract 
constraints related to valid ranges, those are only a small portion of DL-specific constraints (Section~\ref{extraction-results}).  
Techniques such as C2S~\cite{c2s} require pairs of Javadoc comments  and formal JML~\cite{jml} constraints as input. For DL API functions, such formal constraints are unavailable.

\subsection{Our Approach}

To fill this gap, we design and implement a new technique---\emph{\tool}---to analyze API documentation to extract DL-specific input constraints. 
\tool 
features a novel method to automatically derive {\em constraint extraction rules} from a small set of manually annotated API documents with precise constraint information. 
These rules can predict API parameter constraints based on 
syntactic patterns in the form of dependency parse tree
in documents. They are then applied to the full sets of API documents of popular DL libraries to extract constraints.

To demonstrate the effectiveness of these extracted  constraints, \tool  uses 
them
to guide and improve an important task~---~generating test cases automatically
to test DL API functions.
Testing API functions of DL libraries (e.g., TensorFlow~\cite{abadi2016tensorflow} and PyTorch~\cite{NEURIPS2019_9015}) is crucial because these libraries are widely used and
contain software bugs~\cite{zhang2020an, DL_bug_char, tf_bugs, 2020-Humbatova-ICSE, cradle}, which hurt not only the development but also the accuracy and speed of the DL models.

Yet, generating test cases for DL libraries' API functions is challenging.
If a test-generation tool is (1) unaware of DL-specific constraints or (2) incapable of using these constraints to generate diverse inputs, it is practically impossible to generate valid inputs 
to reach deeper states and test the core functionality of DL API functions.
Existing test-generation tools~\cite{afl, honggfuzz, libfuzzer, randoop,TSE12_EvoSuite, FuzzFactory} such as AFL~\cite{afl}
and libFuzzer~\cite{libfuzzer} have no knowledge of such input constraints, thus are very limited in testing DL API functions. 
\tool  addresses these challenges by using  the following  techniques:

\vspace{0.03in}
\noindent\textbf{(1) DL-specific constraint extraction:}
Since API documents are written informally in a natural language, manually extracting constraints from a large number of API documents (e.g., TensorFlow v2.1.0 has 2,334 pages of API documents and 854,900 words) is inefficient and tedious. In addition, since these documents are constantly evolving, it is undesirable and error-prone to manually analyze them each time the documents are updated which can be as frequent as every commit. To address these challenges, we develop a novel method that can 
automatically derive a set of rules 
that predict parameter constraints
from 
parse tree
patterns of API description. Given a small set of 
API function descriptions and the corresponding constraint annotations, \tool identifies a set of rules as an optimal mapping that can minimize prediction errors and achieve the maximum coverage of constraints.
By applying these constructed rules to a much larger set of real-world documents, \tool can
automatically extract
DL-specific constraints for API functions of the most widely used libraries.

\vspace{0.03in}
\noindent\textbf{(2) DL-specific input generation:}
After extracting DL-specific input constraints (e.g., Fig.~\ref{fig:API_example_constraint}),
\tool uses these constraints to guide test generation to produce valid inputs (e.g, Fig.~\ref{fig:API_example_input}),
invalid inputs, and boundary inputs
(such as -\code{MaxInt},
0, and \code{MaxInt} for the constraint of \dtype of \code{int}). 
\tool  evaluates valid inputs by checking if the API runs successfully without failures, e.g., crashes. If a failure occurs with a valid input, the generated test has manifested a bug in the implementation of the API's core functionality.

Fig.~\ref{fig:API_example_all} shows a previously unknown bug detected by \tool in
TensorFlow
along with its patch that the 
TensorFlow 
developers committed after we reported the bug. 
The API document in Fig.~\ref{fig:API_example} indicates that the shape of \code{input} is 5-D, and \code{ksize} is an integer or a list of 1, 3, or 5 integers. 
\tool automatically extracts the constraints in Fig.~\ref{fig:API_example_constraint} and generates the bug-triggering input in Fig.~\ref{fig:API_example_input}.
Detailed constraint formats are explained in Section~\ref{sec:pattern_miner}. 
\tool generates a valid input. Specifically, the parameter \code{input} is a five-dimensional (5-D) tensor as a constant (\code{tf.constant}), where the five pairs of square brackets denote a five-dimensional tensor. Parameter \code{ksize} is a list of length 1, whose element is a zero (i.e., \code{[0]}), the parameter \code{strides} is \code{[1]}, and parameter \code{padding} is \code{"VALID"}.

This bug is only triggered when the parameter \code{ksize} has a zero value. This zero value causes a division-by-zero fault, resulting in a floating point exception. 
To trigger this bug, the parameter \code{padding} must be either \code{"VALID"} or \code{"SAME"}. Otherwise, the function's input validity checking would reject the input with an \code{InvalidArgumentError}. Therefore, it is practically impossible for techniques that randomly generate inputs to
trigger this bug.
After we reported this bug, the TensorFlow developers added the \code{non\_negative} range validation for the parameter \code{ksize} (Fig.~\ref{fig:API_example_fix}).

In addition, 
\tool generates invalid inputs that violate the constraints to 
detect crashes. 
Despite invalid inputs, DL API functions should not crash. 
Instead, they are expected to report an invalid input (e.g., by throwing an exception or printing an error message).
This point is well confirmed by an API developer after we reported a crash bug detected by \tool ``\emph{A segmentation fault is never OK and we should fix it with high priority}”. 
Such invalid input generation is impossible without the constraints.

\vspace{0.03in}
\noindent\textbf{(3) Documentation-bug detection:}
Since incorrect API documentation provides false information about APIs, which often misleads 
API users
to introduce bugs in code~\cite{icomment},  it is  important to detect bugs in API documents as well. 
Different from prior work~\cite{icomment, tcomment} that detects inconsistencies between documents/comments and code, 
\tool detects inconsistencies within documents. 
For example, in the document of \code{tf.keras.backend.moving\_average\_update}, the description for the parameter \code{value} is \textit{``
...with the same shape as \textasciigrave variable\textasciigrave,...''},
but the parameter \code{variable} is not documented.
This documentation bug of erroneous parameter dependencies
has been fixed after we reported it.

\begin{figure*}[htbp]
\centering
\includegraphics[width=0.95\textwidth]{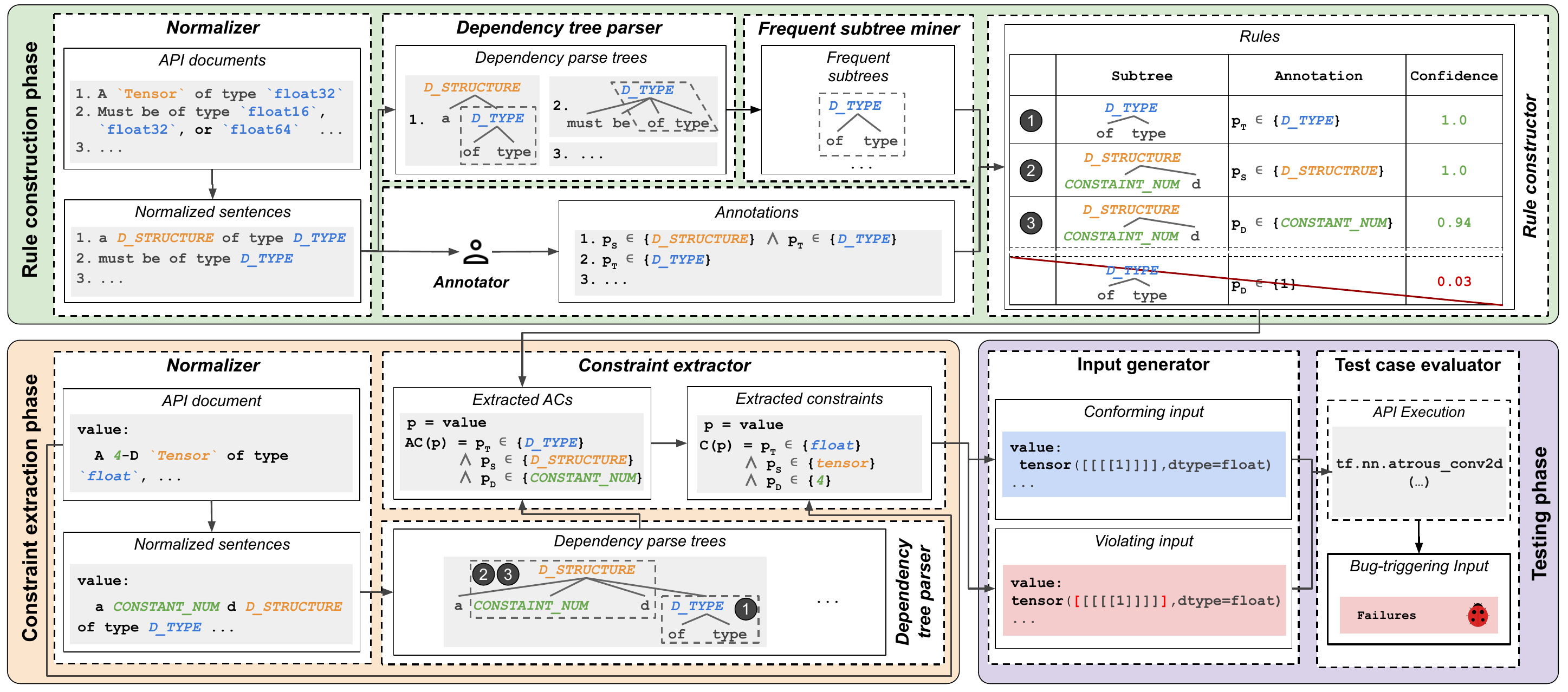}
\caption{Overview of \tool. $p_T$, $p_S$, and $p_D$ are the abstract constraints representing the \textit{data type}, \textit{data structure}, and \textit{number of dimensions} of parameter $p$, respectively. Detailed constraint formats are explained in Section~\ref{sec:pattern_miner}.
}
\label{fig:tech-overview}
\vspace{-4mm}
\end{figure*}

\subsection{Contributions}

In this paper, we make the following contributions:
\begin{list}{\labelitemi}{\topsep=1pt\parsep=1pt\leftmargin=10pt\itemindent=1pt}
 \item A novel rule construction technique that formulates the challenge as an optimization problem aiming to find the smallest set of rules that can make the largest number of correct extractions of parameter constraints. We also develop an approximate solution to the problem based on sample space conditional probability computation. 
\item A document-analysis technique that extracts \totalconstraint~constraints automatically 
from API documentation with the focus on four categories of input properties in DL APIs: \datastructure, \dtype, \shape, and \textit{valid values} for \totalapiwithconstraint~API functions across the three widely-used DL libraries, TensorFlow~\cite{abadi2016tensorflow}, PyTorch~\cite{NEURIPS2019_9015}, and MXNet~\cite{chen2015mxnet}. The constraint extraction precision is \pretotal\%.

\item An application of our extracted constraints to guide the generation of 
DL-specific inputs. 
\item A tool \emph{\tool} that combines the techniques above, and detects  
\oursAll bugs in 
\numAPICrashTotal APIs from
the three libraries, while a baseline  that generates inputs without the knowledge of constraints detects \baseAll bugs only.  
Among the \oursAll bugs,  \oursNewAll are previously unknown bugs, 
\oursNewVeri of which have been fixed (\newBugFixed) or confirmed (\newBugConfirmed) by the developers after we report them.
Notably, \textbf{one of the previously unknown bugs was added to the CVE vulnerability database} for TensorFlow after we reported it. 
In addition, \tool detects \docBugAll documentation bugs, \docBugVeri of which have been fixed (\docBugFixed) or confirmed (\docBugConfirmed) after we report them.
 \end{list}

\vspace{0.03in}
While our rule construction and constraint extraction techniques are general, the constructed rules and extracted constraints  are domain-specific. We focus on testing DL libraries due to their importance and the lack of available constraint-extraction 
techniques for them. We leave the extension to other domains, e.g., classic machine learning libraries such as  scikit-learn~\cite{scikit-learn}, as future work.

\vspace{0.03in}
\noindent\textbf{Availability:} 
We share the tool \tool, bug list, and data in ~\cite{supplementary-material}.

\section{Approach}
\subsection{Overview}
\label{sec:challanges}

Fig.~\ref{fig:tech-overview} shows the overview of \tool using an example of the TensorFlow API \code{tf.nn.atrous\_conv2d}. 
\tool consists of three phases. 
The \emph{rule construction phase} (i.e., the green box in Fig.~\ref{fig:tech-overview}) takes a small portion of API documents with annotations 
to construct a set of rules 
that can extract concrete constraints from API documents. 
The rules are constructed by an optimization-based method.
They are mappings from  document dependency parse trees to the corresponding abstract 
parameter constraints
in the form of assertions (e.g., on \dtype and \shape), which are called 
{\em Abstract Constraints} (ACs).
In the {\em constraint extraction phase} (i.e, the orange box), the rules are  applied to  concrete API documents to derive concrete parameter constraints. 
To demonstrate the effectiveness of these extracted constraints, in the \emph{testing phase} (the purple box), \tool generates test inputs either conforming or violating the constraints (by the \textit{input generator}), and executes the inputs to detect bugs (by the \textit{test case evaluator}), in an iterative fashion.

A major challenge of 
constraint extraction is analyzing free-form API documentation
written in the natural language~\cite{wong2015dase, tcomment, jdoctor}.
We observe that developers have limited ways to express input constraints in natural language. However, these expressions are instantiated differently for different APIs and composed together in various ways, leading to complex overall syntactic structures that are difficult to translate to parameter constraints.
We hence devise a novel method that works as follows. It first preprocesses/normalizes the documents to dependency parse trees and then breaks the trees into subtrees. With a small set of API documents and the corresponding manually annotated ACs,
an algorithm is developed to identify the optimal mappings between subtrees and parameter constraints that can maximize the matching of the mapped constraints and the ground-truth annotations.  
These mappings are essentially our constraint extraction rules. 
\tool applies these rules to  extract constraints from API documents automatically.
Apart from a fixed cost of annotating a small portion (e.g., 30\%) of API parameters, our process is automatic and can be reapplied to future versions or another relevant
library with little manual work.

\vspace{0.03in}
\noindent\textbf{Preprocessing:}
During both the rule construction and constraint extraction phases, \tool performs two preprocessing steps: \emph{normalization} and \emph{dependency tree parsing} to convert free-form API descriptions to dependency parse trees (parse trees for short). (Fig.~\ref{fig:tech-overview}).
The  normalizer replaces keywords with abstractions, e.g., replacing data type keywords (e.g., \code{int32}, \code{float64}) with \code{D\_TYPE}, structure type keywords (e.g., \code{Tensor}, \code{list}) 
with \code{D\_STRUCTURE}, and  integer constants with \code{CONSTANT\_NUM}. This normalization improves the performance of the rule construction algorithm by suppressing instance differences.
We use dependency tree parser~\cite{corenlp} to convert the normalized sentences to parse trees.

\vspace{0.03in}
\noindent\textbf{An example:}
Our rule construction component identifies a rule \encircle{1} (row one of the \textit{Rules} table in Fig.~\ref{fig:tech-overview}) that maps a frequently occurring  subtree pattern ``\textit{of type \code{D\_TYPE}}'' (e.g., appearing in both sentences 1 and 2 in Fig.~\ref{fig:tech-overview}) to an abstract type assertion (an AC) $p_T\in \{$\code{D\_TYPE}{}$\}$,
which means that the valid \dtype of parameter $p$ should be one from the set \{\code{D\_TYPE}{}\}, where \code{D\_TYPE} is an abstraction of one or more \dtypes, which, in this example, are \code{float16}, \code{float32}, and \code{float64}.
In the annotated dataset,
the conditional probability of the type assertion $p_T\in \{$\code{D\_TYPE}{}$\}$, given the subtree pattern is 1.0 and the pattern is the smallest with such predictive power.
Thus, the rule constructor is able to create rule \encircle{1}.

The \emph{constraint extractor} applies 
constructed rules
to all the preprocessed API documents to automatically extract a set of constraints for each input parameter. For example, in the Tensorflow document for API \code{tf.nn.atrous\_conv2d}, one of the parse trees parsed from the description for parameter \code{value} (e.g., Fig.~\ref{fig:tech-overview}) contains two frequent subtrees \textit{\q{a \code{CONSTANT\_NUM} d \code{D\_STRUCTURE}}} and \textit{\q{of type \code{D\_TYPE}}}. These structures correspond to rules \encircle{1}, \encircle{2}, and \encircle{3}. 
\tool applies these rules and obtains the extracted ACs for the parameter in Fig.~\ref{fig:tech-overview} (the middle of the orange box),  
e.g., $p = value,~AC(p) = p_T  \in \{\code{D\_TYPE}{}\} \wedge p_S  \in \{\code{D\_STRUCTURE}{}\} \wedge p_D  \in \{\code{CONSTANT\_NUM}{}\}$
, where $p_T$, $p_S$ and $p_D$ represent the \textit{data type}, \textit{data structure} and \textit{number of dimensions} of parameter $p$, respectively.
\tool further instantiates the abstract symbols (\code{D\_TYPE}, \code{D\_STRUCTURE}, and \code{CONSTANT\_NUM})  with the corresponding value and types (i.e., \code{float}, \code{Tensor}, and 4) from the original sentence to convert the ACs to concrete constraints. 
We now discuss  
each individual step.

\subsection{Preprocessing}
\label{sec:pattern_miner}
The first step of \tool is to collect the natural language API documents. They are at high volume. For example, there are 2,334 pages of API documents and 854,900 words in TensorFlow v2.1.0.
It is hence a daunting and tedious task for developers to manually examine such a large set of API documents to identify constraints.

\vspace{0.03in}
\noindent\textbf{API document collection and tokenization:} 
After collecting the API documents (in the form of HTML pages from DL libraries' websites), \tool parses these files
to obtain API signatures and parameter descriptions
with an HTML parsing tool~\cite{beautifulsoup}.
Since sentence is a natural unit of organizing constraints, \tool further splits the description into sentences with 
a sentence segmentator~\cite{nltk}.

\vspace{0.03in}
\noindent\textbf{Normalization:} The tokenized  sentences are normalized. 
While developers may have a small number of patterns 
expressing parameter constraints, these patterns have
diverse instantiations according to the concrete data
types and parameters involved. Normalization  
abstracts away these instantiation differences.

Specifically, \tool normalizes keywords such as (1) data types (e.g., \code{int32}) and (2) data structures (e.g., \code{tensor}) as \code{D\_TYPE} and \code{D\_STRUCTURE}, respectively. To get the list of keywords for data types, we collect a list of supported data types from each library~\cite{tf_dtype, pt_dtype, mx_dtype}. We then expand such a list with informal variations (e.g., ``integer'', and ``ints'') and missing common types (e.g., \code{String}) to match the format of API documents.
In total, we use 84, 74, and 53 type keywords for TensorFlow, PyTorch, and MXNet, respectively. The full list of keywords can be found in~\cite{supplementary-material}.

\tool also normalizes constants such as (3) integer, (4) float, (5) boolean values as \code{CONSTANT\_NUM}, \code{CONSTANT\_FLOAT}, and \code{CONSTANT\_BOOL}.
It also replaces (6) relational expressions (e.g., ``$\geq 1$'') with \code{REXPR} and replaces (7) parameter names with \code{PARAM}.

The content that is (8) quoted often refers to enumerate values, so \tool replaces such content with \code{ENUM}. For example, \textit{\q{`NWC' and `NCW' are supported.}} is normalized to \textit{\q{\code{ENUM} are supported}}.
The shape of a parameter is often put within (9) a pair of square brackets or parentheses, \tool replaces such content with \code{SHAPE}. For example, \textit{\q{A Tensor of shape [num\_classes,dim]}} is normalized to \textit{\q{A \code{D\_STRUCTURE} of shape \code{SHAPE}}}.
Finally, consecutive abstract annotations of the same type are replaced with just one.
For example, the three type keywords in \textit{\q{Must be of type `float16`, `float32`, or `float64`.}} (Fig.~\ref{fig:tech-overview})  are replaced by a single \code{D\_TYPE}, resulting in a normalized sentence \textit{\q{must be of type \code{D\_TYPE}}}.

\vspace{0.03in}
\noindent\textbf{Dependency tree parsing:}
Once the sentences are normalized, they are fed to the dependency tree parser~\cite{corenlp}, which conducts POS-Tagging and builds tree structure relationships (i.e., dependency parse trees) between words of a sentence based on the grammatical structure. 
For example, in the sentence \textit{\q{a \code{D\_STRUCTURE} of type \code{D\_TYPE}}} from Fig.~\ref{fig:tech-overview}, the words \q{\code{D\_STRUCTURE}}, \textit{\q{type}}, and \q{\code{D\_TYPE}} are first tagged as \code{NN} (noun). Then the parser conducts dependency parsing and generates the dependency parse tree as shown in the figure where \code{D\_STRUCTURE} is the root, and \code{D\_TYPE} is the \textit{nominal modifier}~\cite{dep-standard} of the root.

\vspace{0.03in}
\noindent\textbf{Annotating a subset of API descriptions with ACs:}
To support rule construction, we randomly pick a small set of the parameters (30\%) and manually annotate them with their ACs. To minimize possible biases, the process involves three co-authors. Two authors independently annotate with 98.2\% agreement. All disagreements are resolved with a third author to reach a consensus.

\vspace{0.01in}
\noindent{\em Abstract Constraints (AC):}
ACs are abstract constraints/assertions. These assertions are not on concrete \dtype or \shape but rather abstract ones. An AC for a parameter $p$ is denoted in the form of
$p_t\in \{T_1, T_2,...\}$ where $t$ is the category of AC, and $T_1$ and $T_2$ are the possible abstract values. 
For example, $p_T\in \{\code{D\_TYPE}\}$ means that $p$ is of \code{D\_TYPE}.
Specifically, the annotations of parameter $p$ are designed as follows:
\begin{list}{\labelitemi}{\topsep=1pt\parsep=1pt\leftmargin=15pt\itemindent=1pt}
    \item $p_T$ denotes the \textit{data type} of an abstract constraint (AC) of $p$.
    \item $p_S$ denotes the \textit{data structure} of an AC  of $p$.
    \item  $p_{SP} \in \mathbf{N}^{D}$  where $ D \in \mathbf{N}$ denotes  the \shape AC of $p$, where $D$ represents the number of dimensions of $p$. 
    \item $p_{D} \in \mathbf{N}$ denotes the \textit{number of dimensions} of an AC of $p$. 
    Therefore, $p_{D} = p_{SP}.length$ if $p$ is a tensor or $p_D=0$ if $p$ is a scalar.
    \item  $p_i$ denotes an element in parameter $p$ if $p$ is a tensor, where $i=1,2,...,Prod(p_{SP})$. When $p$ is a scalar, its value is $p_0$. 
\end{list}

An AC can be instantiated to different concrete constraints.
Table~\ref{fig:subtree_example} 
provides examples of ACs (first column as part of the rules) and their instantiations (last column) for several APIs.

\vspace{0.01in}
\noindent{\em AC annotation categories:}
We focus on annotating four categories of ACs (i.e., \textit{\datastructure}, \textit{\dtype}, \shape, and \validvalue)
because they represent the most common (93.6\%)
properties of input parameters of API functions
in major DL libraries.
The four categories are:
\begin{list}{\labelitemi}{\topsep=1pt\parsep=1pt\leftmargin=15pt\itemindent=1pt}
    \item \textit{\datastructure}: the type of data structure that stores a collection of values for the input parameter, 
    such as list, tuple, and n-dimensional array (i.e., tensor). 
    
    \item \textit{\dtype}: the data type, such as \code{int}, \code{float}, \code{boolean}, and \code{String}, of the parameter or the elements of \datastructure.
    
    \item \shape: the shape or number of dimensions of the parameter. 
    For example, in row 2 of Table~\ref{fig:subtree_example}, \code{weights} is of shape
    \code{[num\_classes, dim]} (i.e., it is a 2-D array where the sizes of its first and second dimensions are \code{num\_classes} 
    and \code{dim}, respectively).
    \item \validvalue: a set of valid values (e.g., parameter \code{padding} can only be either \code{"VALID"} or \code{"SAME"})
    or the valid range of a numerical parameter 
    (e.g., a float between 0 and 1).
\end{list}

We make 
reasonable assumptions when annotating API descriptions. For example, a parameter is assumed to be a 0-dimensional non-negative \code{integer} if the document states it is a \textit{``number of ..."}. 
\emph{The assumptions are in the supplementary  material~\cite{supplementary-material}.}

\subsection{Rule Construction}
\label{sec:rule_construction}
Although API descriptions are in a natural language, these descriptions often share a small number of syntactic patterns. For example, a constraint of \dtype assertion is mostly described by two syntactic patterns: \textit{\q{must be one of the following types ...}} and \textit{\q{a tensor of type ...}} in TensorFlow. 
Our idea is hence to identify such patterns in API descriptions and project them to the corresponding parameter constraints.  We call such projections the {\em constraint extraction rules}.

Automatically deriving such rules is challenging. The first challenge is that a syntactic pattern may have  different instantiations in various API descriptions, depending on the variables and types. 
For example, the aforementioned pattern \textit{\q{A `Tensor' of type ...}} is instantiated to \textit{\q{A `Tensor' of type `string'}} and \textit{\q{A `Tensor' of type `int32'}} in two respective parameters \code{contents} and \code{crop\_window} in API \code{tf.io.decode\_and\_crop\_jpeg}. Our normalization step substantially mitigates this problem. The second challenge is that such patterns are often convoluted in the overall syntactic structure of an API description. For example, 
consider the description of parameter \code{value} as shown in Fig.~\ref{fig:tech-overview}. The normalized sentence \textit{\q{a \code{CONSTANT\_NUM} d \code{D\_STRUCTURE} of type \code{D\_TYPE}.}} is composed of two syntactic patterns \textit{\q{\code{CONSTANT\_NUM} d \code{D\_STRUCTURE}}} and \textit{\q{of type \code{D\_TYPE}}}.
Third, these patterns may have arbitrary sizes.  

\vspace{0.03in}
\noindent\textbf{An optimization problem:} 
We propose a novel method to automatically derive the extraction rules from a small set of APIs with their ACs 
manually annotated.  We formulate it as an optimization problem.
Specifically, given an API $f$,  its normalized natural language description is denoted as $D_f$, its ACs 
are denoted as $A_f$.
We use $\mathit{trees}(D_f)$ to denote all the subtrees of the parse tree of $D_f$. For example, Fig.~\ref{fig:tech-overview} gives a 3-layer parse tree of the normalized sentence \textit{\q{a \code{D\_STRUCTURE} of type \code{D\_TYPE}}}, which has subtrees \textit{\q{of type \code{D\_TYPE}}} and \textit{\q{a \code{D\_STRUCTURE} type}}.  
Such subtrees consider both parent-child (direct) connections and ancestor-descendant (indirect) connections. 
We use $\mathit{Tree}$ and $\mathit{AC}$ to denote the domains of subtrees and ACs, respectively. Our goal is hence to derive a mapping $R:\mathit{Tree}\rightarrow \mathit{AC}$.
The mapping should satisfy the following optimization objectives. 
First, the tree patterns can be used to precisely predict the corresponding ACs.
If we consider the tree patterns and the ACs form a distribution, the conditional probability of an AC given the condition of its tree pattern shall be high.
Second, all ACs can be predicted by these patterns, i.e., our mappings should be comprehensive.
Third, the number of the mappings from a tree pattern to an AC in $R$
is minimum.
The objective is needed otherwise a simple solution would be to include all tree patterns in descriptions.
Fourth, the size of each tree pattern is minimal. It is very likely that a tree pattern and its sub-patterns both can predict a constraint. In such cases, we prefer the smallest one, which provides the maximum generalization.  

Formally, the process to find the set of rules, that is, the optimal mappings $R$, is the following. 
\begin{equation}
    \argmin{R}{\E_{(D_f,A_f)\sim\mathcal{N}}}\left[\begin{split}
    &\frac{\sum\limits_{a\in A_f, t\in \mathit{trees}(D_f), R(t)=a}1-P(a|t)}{|A_f|}\ +  \\
    & \ \ \ \ \ |\{a \in A_f| \not\exists t \in \mathit{trees}(D_f)\ s.t.\ R(t)=a\}|\  + \\
    & \ \ \ \ \  |R|\ +  \sum\limits_{a\in A_f, t\in \mathit{trees}(D_f), R(t)=a}|t|\ 
    \end{split}
    \right]
\end{equation}

Here, $\mathcal{N}$ denotes the distribution of API description and the corresponding ground-truth ACs. The above formula means that we are looking for an $R$ that can minimize the expected objective function value for all samples $(D_f,A_f)\sim \mathcal{N}$. The objective function 
is the sum of four terms. The first one is the average conditional probabilities for all the ACs in $A_f$. Intuitively, it means that
for each AC $a$ in an API, our mapping $R$ should associate a tree pattern $t$ with $a$ such that the 
conditional probability $P(a|t)$ is maximum.
The second term means that the number of ACs in $A_f$ for which $R$ does not have a mapping is minimum. This is to maximize the coverage of our rules. The third term is to minimize the size of $R$.
The last term is to minimize the size of each tree pattern in $R$.

\vspace{0.03in}
\noindent
{\bf Approximate solution:} Solving the above optimization problem is difficult because it is discrete. Its complexity is NP. 
This is not a typical learning problem as it does not aim to learn a distribution but rather to construct a minimum and yet complete set of rules.
In addition, the amount of data available for training is relatively small compared to other domains that have successful applications of deep learning models. 
We hence devise an approximate solution. The first term can be approximated by computing the sample space conditional probabilities and then including the top associations in $R$. The sample space conditional probabilities are: 
\begin{equation}
\label{eq:cond_prob}
    \overline{P}(a|t)=\frac{|\ \{f\ |\ a\in A_f \wedge t\in \mathit{trees}(D_f)\ \}\ |}{|\ \{f\ |\ t\in \mathit{trees}(D_f)\}|} 
\end{equation}

\begin{algorithm}
\small
\caption{Rule Construction}\label{euclid}
\label{algo:rule_construction}
\begin{algorithmic}[1]
\Function{RuleConstruction}{$\textit{Sample}, \textit{min\_support}, \textit{min\_confidence}$}
\State $\textit{parser} \gets \Call{DependencyTreeParser}$
\State $D_f, FreqD_f \gets \emptyset  $
\State $R \gets \emptyset$
\Foreach{$\textit{sentence} \in \textit{Sample}$}
    \State $\textit{dependencyParseTree} \gets \textit{parser}.\Call{parse}{\textit{sentence}}$
    \State $D_f.\Call{add}{\textit{dependencyParseTree}}$
\EndForeach

\State $FreqD_f \gets \Call{getFreqSubtree}{D_f,\textit{min\_support},\textit{MAX\_SIZE}}$

\Foreach{$t \in FreqD_f$}
    \Foreach{$a \in \Call{selectAC}{\textit{Sample}, t, \textit{min\_confidence}}$}
        \State $R.\Call{add}{t \to a}$
    
    \EndForeach

\EndForeach

\State \Return $R$

\EndFunction
\end{algorithmic}

\end{algorithm}

Intuitively, it is the number of co-occurrences of a tree pattern and an AC divided by the number of occurrences of the tree pattern. We further observe that if
a tree pattern $t$ is rare, it is usually not related to parameters. As such, we can focus on the frequent subtrees. 
We use frequent subtree mining~\cite{subtreemining} to efficiently discover the most frequent subtrees. We filter out the rare tree patterns with threshold \code{min\_support}, i.e., any tree patterns that occur less than or equal to \code{min\_support} times are discarded.
The second and third terms are approximated by selecting only the associations (of $a$ and $t$) with a large sample space conditional probability.
Specifically, \tool includes in $R$ associations (of $a$ and $t$) with $\overline{P}(a|t)$ 
greater than or equal to
\code{min\_confidence} (the selection of and \code{min\_support} and \code{min\_confidence} is a trade-off between precision and recall and will be discussed in Section~\ref{sec:setup}).
An alternative to approximate the second term is to use a greedy algorithm to include additionally needed tree patterns to achieve (full) coverage of ACs. However, this often contradicts the third term. Empirically (see Section~\ref{extraction-results}), we find that including the top associations provides a good balance.
To approximate the fourth term, which minimizes the tree patterns, we keep only the smallest tree pattern when multiple patterns can be used to predict an AC.

Algorithm~\ref{algo:rule_construction} formalizes the process of finding the approximated solution.
For each sentence in the annotated data (\code{Sample}), the dependency tree parser parses the sentence and generates the parse tree, which is added to the set $D_f$ (lines 5--7). Then, we select the set of frequent tree patterns $FreqD_f$ whose frequencies are at least \code{min\_support} using frequent subtree mining (\textproc{getFreqSubtree}) (line 9). This process keeps only tree patterns whose size is smaller than or equal to \code{MAX\_SIZE}.  
For each frequent tree pattern $t$, \textproc{selectAC} selects a set of ACs ($a$) with probabilities $\overline{P}(a|t)$  
greater than or equal to
\code{min\_confidence}. Then, each association of $a$ and $t$ is added to the set $R$ (lines 10--14).

Table~\ref{fig:subtree_example} shows examples of the automatically discovered rules by \tool (col. ``Extraction rules'') and examples of matched sentences (col. ``API sentences'').
For example, rule \encircle{3} in Table~\ref{fig:subtree_example} is used to extract the enumerated value (e.g., \validvalue) of parameter $p$, which is associated with the AC $p_0 \in {ENUM}$. 
In rule \encircle{4}, the pattern \textit{\q{number of}} implies 
parameter \code{p} 
should be a 0-dimensional non-negative integer.

\begin{table*}[t]
	\centering
	\caption{Rule examples and the extracted constraints from TensorFlow, PyTorch, and MXNet
	}
	\label{fig:subtree_example}
	\vspace{-4mm}
	\includegraphics[width=0.85\linewidth]{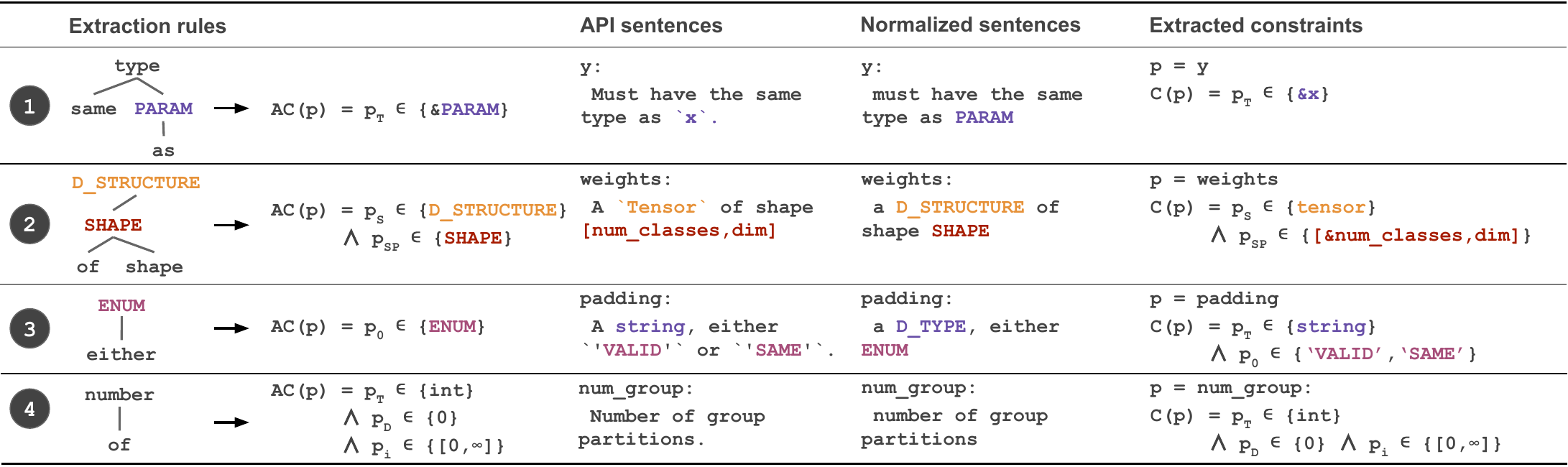}
	\vspace{-3mm}
\end{table*}
\subsection{Constraint Extraction}
\label{approach_constr_extractor}

Given an API description, the constraint extractor matches the tree patterns in the rules with the parse tree of the description. Matches are then projected to the corresponding ACs, which are further instantiated in the context of the description to derive the concrete constraints. 
For example, in Fig.~\ref{fig:tech-overview}, the constraint extractor finds rules \encircle{1}, \encircle{2}, and \encircle{3}  match the two subtrees \q{a \code{CONSTANT\_NUM} d \code{D\_STRUCTURE}} and \q{of type \code{D\_TYPE}} in the parse tree of the \code{value} description. 
\tool then assigns the three relevant ACs to the parameter \code{value}. \tool then instantiates the ACs with the concrete data types, structure types, and constants to generate the final constraints. In row 2 of Table.~\ref{fig:subtree_example}, the annotation \code{SHAPE} is instantiated based on the original text, i.e., \textit{``[num\_classes, dim]''}. Row 3 in Table.~\ref{fig:subtree_example} shows an example rule of \validvalue constraints. \tool detects the pattern \textit{``either ENUM''} and uses it to extract the \validvalue constraint in the last column.

\vspace{0.03in}
\noindent\textbf{Constraint dependencies graphs:}
\label{sec:constr_dep}
The description of one parameter often refers to the 
\dtype or value
of another
parameter 
from the same API function. 
In such cases,
\tool extracts constraints that involve \emph{dependencies} among input parameters.
These 
dependencies are useful not only for generating valid inputs but also for determining the  parameters' generation order.
The automatically constructed rules can detect \dtype dependencies. For example, row 1 in the Table.~\ref{fig:subtree_example}) shows the pattern \textit{``must have the same type as ...''} which indicates a type dependency. 
In this example, the operator `\&' is to acquire the \dtype of a parameter.
An example of \shape dependency is shown in row 2 of Table~\ref{fig:subtree_example}.
Specifically, parameter \code{weights} should have shape \code{[num\_classes,dim]} where the size of the first dimension is the value of
another 
parameter \code{num\_classes} while  \code{dim} is a non-negative integer.

These dependencies are denoted in a graph with each edge representing a constraint dependency. During input generation, the graph is traversed in a topological order to ensure dependencies are properly considered. The graph construction is straightforward and hence elided.

\vspace{-2mm}
\subsection{Testing Phase}
\label{sec:fuzzing_process}

To demonstrate the effectiveness of the extracted constraints, for each API function, \tool iteratively generates an input i.e., values of the API function's parameters,
and evaluates that input to detect crashes.
By either following or violating the
\emph{extracted} constraints, the input generator generates \textit{conforming inputs} (CIs) or \textit{violating inputs} (VIs),  respectively.
The conforming inputs are designed to test the core functionality of the API function while the violating inputs 
aim
to test the API function's input validity checking code.
In both cases, \tool reports bug-triggering inputs that cause serious crashes (e.g., segmentation fault).
\tool tests each API function with \code{maxIter} number of inputs, and the ratio of inputs allocated to each mode (CI or VI) is determined by the ratio \code{conform\_ratio}.

\vspace{0.03in}
\noindent\textbf{Input generator:} 
In each iteration, \tool generates values for all required parameters and some optional parameters (for testing more diverse code).  
The probability for generating each optional parameter 
is \code{optional\_ratio}.

The input generator generates one input for each 
iteration. Given the extracted constraints, \tool generates a value for each parameter following the  order determined by  constraint dependencies (Section~\ref{approach_constr_extractor}). For a conforming input, all generated arguments satisfy the extracted constraints for 
\datastructure, \dtype, \shape, and \validvalue.
If concrete values are specified (e.g., enumerated values) in the constraints, the input generator chooses from those values.
Otherwise, it chooses a \dtype from the list of \dtypes specified in the 
constraints and creates a \shape following the constraints. If the constraints do not specify valid \dtypes, 
\tool selects one from a default list of \dtype described in Section~\ref{sec:pattern_miner}. 
While the input generator is choosing \dtype and \shape for a parameter, it
ensures they are generated according to the parameter dependencies, if any.
For example, parameters often have matching dimension(s), so the input generator needs to ensure such shape consistency.

Once the \dtype and \shape are determined, the input generator generates an n-dimensional array
with values satisfying the given \dtype, \shape, and the range as specified in the constraints, if any.
Finally, the \datastructure constraints are checked and satisfied. For example, if the generated value is 1-dimensional and the constraints explicitly specify the \datastructure (e.g., tuple or list) for the parameter, the input generator converts the generated value accordingly.

To generate an invalid input,  
the input generator randomly selects one parameter and generates a value that violates one or multiple relevant constraints of that  parameter. For all other parameters, \tool generates their values in the same way as conforming inputs (i.e., conforming to all constraints).

\vspace{0.03in}
\noindent\textbf{Constraint-guided boundary-input generation:}
Boundary input values (e.g., 0 and \code{None}) tend to cause bugs due to off-by-one errors etc.~\cite{Audee, probfuzz}. Thanks to the extracted constraints, 
\tool  generates boundary values that follow the constraints and boundary values that violate the constraints.
For each API, \tool picks one parameter with the probability of \code{mutation\_p} to be mutated to one of the boundary cases. We consider six types of boundary mutators: one constraint-specific (boundary values of constraints) and five generics (\code{None}, zero,  zero dimension, empty list, and empty string). As an example, the mutator ``zero dimension" sets the size of one of the dimensions of the parameter's shape to 0 (e.g., it mutates a 3-D tensor of shape \code{[1,1,1]} to \code{[1,0,1]}).

\vspace{0.03in}
\noindent\textbf{Test case evaluator:}
\label{test-eval}
The test case evaluator invokes the target function with the generated input.
If a severe failure occurs,
\tool reports the input as a bug-triggering input.
Specifically, \tool 
returns those inputs causing 
a segmentation fault, floating-point exception, abort, and bus error 
in the C++ backend. We exclude 
aborts caused by assertion failures in MXNet since MXNet uses those for exceptions.   
Crashes from the C++ backend (which handles computationally-intensive DL tasks)
indicate severe problems.

\label{sec:approach}

\section{Experimental Setup}
\label{sec:setup}

\noindent\textbf{Data collection:}
We choose three popular DL libraries (TensorFlow 2.1.0, PyTorch 1.5.0, and MXNet 1.6.0)
as test subjects.
There are 144,541--854,900
words in the collected API documents. 
Among them, we consider 
1008, 529, and 1021 relevant APIs for the three respective libraries.
An API is 
irrelevant if it
(1) is deprecated, (2) has no input argument, (3) cannot be  parsed due to HTML syntactic errors and typos, 
(4) is a non-layer class constructor,  or (5) has an API document without a ``Parameter" description section. 
In total, 2,666 APIs are filtered out due to the five reasons above, 53.5\% of which are due to reason (1). We list the break down in~\cite{supplementary-material}.

\vspace{0.03in}
\noindent\textbf{Rule construction and constraints extraction: } 
\tool applies three thresholds \code{MAX\_SIZE}, \code{min\_support}, and \code{min\_confidence} to construct extraction rules (Section~\ref{sec:rule_construction}). We set \code{MAX\_SIZE} to 7 to all three libraries.
To select the best value for 
\code{min\_support}and \code{min\_confidence}, 
we conducted 5-fold cross-valida\-tion on the 30\% annotated data and measure the quality of the extracted constraints.
By selecting the best F1 score, we set \code{min\_support} to 10, 10, 20, and \code{min\_confidence} to 0.9, 0.7, 0.9 for TensorFlow, PyTorch, and MXNet, respectively. 

\vspace{0.03in}
\noindent\textbf{Input generation and testing: }
We use Docker with Ubuntu 18.04, TensorFlow 2.1.0, PyTorch 1.5.0, and  MXNet 1.6.0.
For multi-dimensional arrays,
\tool generates shapes of 0--5 dimensions or as specified by the constraints. 
By trying different values of \texttt{optional\_ratio} and \texttt{mutation\_p}
on 10\% randomly sampled APIs, we choose \texttt{optional\_ratio=0.2} and \texttt{mutation\_p=0.4}.

\vspace{0.03in}
\noindent\textbf{Manual and execution time:} The AC annotation~(Section~\ref{sec:pattern_miner}) takes \manhour manual hours. 
\tool takes 
34 minutes to perform 
rule construction and constraints extraction for all 
libraries. 
On average, it takes \tool 0.14 seconds to generate and test each input.

\section{Evaluation and Results}

\label{eval}

We
answer four research questions (RQs): 
\textbf{RQ1: }How effective is \tool in extracting constraints from DL API documentation? (Section 4.1) 
\textbf{RQ2: }How is \tool compared to existing constraint-extraction approaches? (Section 4.2) 
\textbf{RQ3: }Can the extracted constraints enable \tool to detect more bugs? (Section 4.3) and  
\textbf{RQ4: }How effective is \tool in generating valid inputs? (Section 4.4)

\subsection{RQ1: Effectiveness of Constraint Extraction} 
\label{extraction-results}

\noindent\textbf{Approach: }
We apply \tool to extract constraints in our subjects and study the number and quality of constraints. 
We randomly sample an extra 5\% (\tolnumdoc) of input parameters (excluding the 30\% AC annotated data for rule construction) to form an evaluation set.
We manually annotate these parameters with concrete constraints
to build the ground truth.
The constraints extracted by \tool are then compared against the evaluation set.

For each parameter, we consider all valid options for one category as one constraint. And the constraint for this category is correct iff all valid options are correctly extracted. 
For example, the parameter \code{size} of \code{tf.slice} 
can be either \code{int32} or \code{int64}. The extracted \dtype constraint $p_T \in \code{\{int32,int64\}}$
is deemed correct, while
$p_T \in \code{\{int32\}}$
is considered incorrect. 
If a parameter's document contains no constraints of the four categories,
it is excluded from the 
precision and recall computation. 
While it is reasonable to include such no-constraint parameters in our calculation because \tool can trivially extract nothing, the accuracy may be inflated if there is a large portion of such parameters.
Among the sampled parameters, the numbers of no-constraint parameters
are 29 (15.3\%), 10 (10.8\%), and 91 (28.4\%) for TensorFlow, PyTorch, and MXNet, respectively (details in Extraction result section below).

We use the standard metrics precision, recall, and F1 score of the extracted constraints of the sampled parameters for each constraint category. 
\emph{Precision} is the percentage of the correctly extracted constraints (i.e., extracted constraints that match the ground-truth) over the number of all extracted constraints.
\emph{Recall} is the percentage of correctly extracted constraints over the total number of all ground truth constraints. \emph{F1} is the harmonic mean of precision and recall.

\begin{table}[t]
\centering
\caption{Quality of constraint extraction }
\vspace{-3mm}
\label{tab:constraint_quality}

\resizebox{\linewidth}{!}{
\begin{tabular}{@{}l@{ }rrrr@{}}
\toprule
& \textbf{TensorFlow} & \textbf{PyTorch} & \textbf{MXNet} & \textbf{Total/Avg}\\

\midrule
\# APIs with  constr. extracted &   \tfapiwithconstraint           &\pytorchapiwithconstraint           &    \mxnetapiwithconstraint     & \totalapiwithconstraint  \\

\# constr. extracted           & \tftotalconstraint               &         \pytorchtotalconstraint        &      \mxnettotalconstraint    & \textbf{\totalconstraint}     \\

\# constr. per API: \textit{Avg (Min-Max)}                            & 6.5 (1-51)                   &     6.4 (1-33)             &     6.9 (1-111)     &   6.6 (1-65)    \\

\midrule

\# evaluated param.    & \tfnumdoc           &     \pytorchnumdoc          &       \mxnetnumdoc         & \tolnumdoc        \\

\# evaluated param. with constr. &161 & 83 & 229 & 473\\

\# evaluated constr.   & \tfnumconstraint    &     \pytorchnumconstraint   &       \mxnetnumconstraint  & \tolnumconstraint \\

\midrule
Precision/Recall/F1 for All (\%)                 &    \tfpretotal/\tfrecalltotal/\tffscoretotal      &      \ptpretotal/\ptrecalltotal/\ptfscoretotal        &       \mxpretotal/\mxrecalltotal/\mxfscoretotal    &   
\textbf{\pretotal}/\recalltotal/\fscoretotal\\

\midrule

Precision/Recall/F1 for \dtype (\%)                  &    \tfpredtype/\tfrecalldtype/\tffscoredtype      &      \ptpredtype/\ptrecalldtype/\ptfscoredtype     &       \mxpredtype/\mxrecalldtype/\mxfscoredtype        &  
\predtype/\recalldtype/\fscoredtype\\

Precision/Recall/F1 for \datastructure (\%)                   
&    \tfpreds/\tfrecallds/\tffscoreds      
&      \ptpreds/\ptrecallds/\ptfscoreds         
&       \mxpreds/\mxrecallds/\mxfscoreds           
&  \preds/\recallds/\fscoreds \\

Precision/Recall/F1 for \shape (\%)                  
&    \tfpreshape/\tfrecallshape/\tffscoreshape   &      \ptpreshape/\ptrecallshape/\ptfscoreshape       &      \mxpreshape/\mxrecallshape/\mxfscoreshape        &   
\preshape/\recallshape/\fscoreshape \\

Precision/Recall/F1 for \validvalue (\%)                 &    \tfprevalidvalue/\tfrecallvalidvalue/\tffscorevalidvalue      & \ptprevalidvalue/\ptrecallvalidvalue/\ptfscorevalidvalue &      \mxprevalidvalue/\mxrecallvalidvalue/\mxfscorevalidvalue         & \prevalidvalue/\recallvalidvalue/\fscorevalidvalue   \\

\bottomrule

\end{tabular}
}
\vspace{-5mm}
\end{table}

\vspace{0.03in}
\noindent\textbf{Extraction results: }
Table~\ref{tab:constraint_quality} shows
the quality of extracted constraints.
In total, \tool extracts \totalconstraint constraints automatically from the three libraries (row \textit{\#constr. extracted}).
Specifically, 
TreeMiner~\cite{subtreemining} collects 873, 426, and 321 frequent parse subtrees from the three libraries respectively with the corresponding \code{min\_support}. Then \tool constructs rules with \tfnumpat, \pytorchnumpat, and \mxnetnumpat subtrees. 
The remaining subtrees do not constitute any rules because no
AC is associated with the subtree with large enough conditional probability.
Using these rules, \tool extracts on average 6.6 constraints per API for all three libraries (row \textit{\#constr. per API: Avg (Min-Max) Table~\ref{tab:constraint_quality}).}
Overall, \tool can
extract constraints from   
90.4\%, 94.1\%, and 98.5\%  of relevant APIs (details in Section~\ref{sec:setup}) for TensorFlow, PyTorch, and MXNet, respectively.

For each library, Table~\ref{tab:constraint_quality} shows the number of parameters in the evaluation set~(row \textit{\#evaluated param.}), the number of the parameters in the evaluation set with at least one constraint~(row \textit{\#evaluated param. with constr.}), the number of constraints manually labeled in the evaluation set~(row \textit{\#evaluated constr.}).
The \textit{Total/Avg} column shows the total number of parameters and constraints in the evaluation set, and the average precision, recall, and F1 score.

Overall, \tool achieves a high precision~(\pretotal\%) and recall~(\recalltotal\%) of constraint extraction across all three subjects.
\tool is quite effective in extracting constraints for \dtype and \datastructure with F1 score over 80\%. It is less effective when extracting constraints for \validvalue. 
The reason is that sentences that describe constraints for \validvalue are 
not as common in the annotated data compared with other categories, e.g., \datastructure, and thus \tool misses some  patterns given the \code{min\_support} and \code{min\_confidence} thresholds.
For example, when analyzing the sentence \q{\textit{Only `zeros' is supported for quantized convolution at the moment}}, \tool misses the \validvalue constraint $p_0\in\{\code{`zeros'}\}$ because the pattern \q{\textit{Only ... is/are supported}}  is not  frequent enough in the annotated dataset and did not pass the set thresholds.
In Fig.~\ref{fig:API_example_all}, \tool misses the constraint that parameter \code{ksize} can also be a single integer due to the same reason.
With the extracted incomplete constraints (i.e., \code{ksize} is a list of integers), \tool still detects the bug. This confirms that to detect real-world bugs effectively, one does not need to have complete constraints~\cite{wong2015dase}.
We choose the thresholds as a trade-off between precision and recall, and one can choose lower thresholds for a better recall.

To show the impact of our rule extraction design (Section~\ref{sec:rule_construction}), we conduct an ablation study, where we do not use the conditional probabilities (Eq.~\ref{eq:cond_prob}) when constructing rules (the set $R$) and instead set the \code{min\_confidence} to 0 (instead of the settings in Section~\ref{sec:setup}). Under these settings, any frequent subtree in $trees(D_f)$ and any AC in $A_f$ that has more than one co-occurrence will be considered as a rule. As a result, this version of \tool without the conditional probabilities extracts 1,621 imprecise rules and extracts constraints with an F-1 of 27.9\% on the same evaluation set (Table~\ref{tab:constraint_quality}). 
In contrast, our \tool, with the conditional probabilities, constructs \totalnumpat rules~(Table~\ref{tab:constraint_stat}) and extracts constraints with a much higher F-1 of \fscoretotal\% (Table~\ref{tab:constraint_quality}) on the same data.

Overall, \tool extracts tens of thousands of correct constraints for these libraries, which enables the generation of valid inputs for  detecting \oursAll bugs.
We show the breakdown of the number of rules and constraints extracted for all three libraries in Table~\ref{tab:constraint_stat}. Note that a subtree can be mapped to multiple categories of ACs.

\vspace{0.03in}
\noindent\textbf{Sensitivity study:} Since one can choose to annotate fewer parameters to save manual effort at the cost of a reduced F-1 score, we quantify the trade-off between the effectiveness of our approach (measured by F-1) and manual effort, which is measured by the number of parameters to annotate. 
Specifically, we evaluate the F-1 scores of our approach by using different amounts of the annotated data, i.e., 5\%, 10\%, and 30\% of parameters. The 5\% of parameters are a subset of the 10\% of the parameters, which is a subset of the 30\% of the parameters.  
\tool  achieves an overall F-1 score of 66.0\% with just 514 annotated parameters (5\% of the parameters), 73.9\% with 1,028 annotated parameters (10\% of parameters), and \fscoretotal\% with 3,086 annotated parameters (30\% of the parameters).

\begin{table}[t]
\centering
\caption{Number of rules and constraints
} 
\label{tab:constraint_stat}

\resizebox{1\linewidth}{!}{
\begin{tabular}{@{}l@{}r@{ }rr@{ }rr@{ }rr@{ }r@{}}
\toprule
\multirow{2}{*}{\textbf{Category}} &
  \multicolumn{2}{c}{\textbf{TensorFlow}} &
  \multicolumn{2}{c}{\textbf{PyTorch}} &
  \multicolumn{2}{c}{\textbf{MXNet}} &
   \multicolumn{2}{c}{\textbf{Total}}\\
  \cmidrule(rr){2-3} \cmidrule(lr){4-5} \cmidrule(ll){6-7} \cmidrule(ll){8-9} 
  
  & Rules & Constraints & Rules & Constraints & Rules & Constraints  & Rules & Constraints\\ \midrule
  \textbf{\dtype} & 405 &  2,392& 114 & 1,163 &196  &  2,272 & 715 & 5,827 \\
  \textbf{\datastructure} &230  & 1,305  &151  &890 & 78 &2,466  & 459 & 4,661  \\
  \textbf{\shape}      &306  &  1,825& 282 & 852 & 173 & 1,699 & 761 & 4,376  \\
  \textbf{\validvalue}   & 97 & 386 & 22 &296  & 11 &489   & 130 & 1,171 \\ \midrule
  \textbf{All}    &\tfnumpat  &\tftotalconstraint  & \pytorchnumpat & \pytorchtotalconstraint &\mxnetnumpat & \mxnettotalconstraint  & \totalnumpat & \totalconstraint  \\ 
  
  \bottomrule
\end{tabular}
}
\vspace{-3mm}
\end{table}

\vspace{0.03in}
\noindent\textbf{Generality of rules: }
To evaluate the generality \emph{across libraries}, we apply the rules constructed from TensorFlow and MxNet to the documentation of PyTorch, and get the constraints with precision, recall, and F1 of 87.9\%, 70.3\%, and 78.1\%. In addition, with the rules \tool constructed from all three libraries, we extract 2,312 constraints from 223 scikit-learn APIs’ documents. 
We manually inspect the extracted constraints on 5\% (59) randomly sampled parameters of scikit-learn APIs. \tool achieves a precision/recall/F1 of 71.3/66.1/68.6\%.
The results suggest that \emph{\tool can be applied to new libraries completely automatically without requiring annotating any documents of the new libraries.}  

To evaluate the generality \emph{across versions} of the same library,
we apply the rules that \tool constructed from TensorFlow v2.1.0, PyTorch v1.5.0, and MXNet v1.6.0 to six versions: two more recent versions from each library respectively (TensorFlow  v2.2.0 and v2.3.0, PyTorch v1.6.0 and v1.7.0, and MXNet v1.7.0 and 1.8.0). \tool extracts  59,936 constraints from 7,684 APIs containing 580,187 words. For evaluation, we randomly sample 603 parameters (same number of parameters as the evaluation in Table~\ref{tab:constraint_quality}) that are either with updated descriptions or newly added to the more recent versions. 
The results show that the rules that \tool constructed are general across versions and can extract constraints from other versions with a precision/recall/F1 of \prevereval/\recallvereval/\fscorevereval\%.

\subsection{RQ2: Comparison with Existing Approaches}
\label{sec:comparison}

We compare \tool with grep and state-of-the-art constraint extraction approaches (e.g., Jdoctor~\cite{jdoctor}). 

\vspace{0.03in}
\noindent\textbf{Comparison with grep:}  
While it may appear to be straightforward to use a \textit{grep-like} technique (i.e., matching existing keywords in documents) to extract constraints, 
such a technique can only identify relevant API document sentences.
\tool, on the other hand, extracts concrete constraints automatically.
The \textit{grep-like} approach could assign a constraint to a match, 
e.g., if a sentence contains the keyword ``integer'', the corresponding parameter would be assigned the constraint $p_T\in\{\code{int}\}$.
We implement this approach by
searching in the documents for keywords
of \dtype constraints~(e.g., ``int'' and ``integer'')
, and \datastructure constraints~(e.g., ``list'' and ``tensor'').
We manually collect such keywords in 
the API documents.
This 
approach misses 47.8\%
of the constraints that \tool extracts, 
i.e,  all \textit{shape}, all \textit{valid value}, 33\%
of \dtype, and 4\% of \datastructure constraints.

\vspace{0.03in}
\noindent\textbf{Comparison with existing constraint-extraction techniques:}
We compare \tool with the state-of-the-art constraint-extraction techniques, including Jdocter~\cite{jdoctor}, DASE~\cite{wong2015dase}, Zhou et al.~\cite{zhou2017analyzing}, and Advance~\cite{RTFM-advance}.
We exclude C2S~\cite{c2s} because it requires 
 formal specifications (JML~\cite{jml}), which is unavailable for the three libraries. 
We exclude Pytype~\cite{pytype} because it cannot analyze across Python and C++ code, therefore, cannot precisely infer types for DL libraries.
Jdocter~\cite{jdoctor}, DASE~\cite{wong2015dase}, and Advance~\cite{RTFM-advance} can only extract constraints for \validvalue (e.g., range). 
With the assumption that they can extract all \validvalue constraints correctly, the best (upper bound) recall that these tools can achieve is 11.9\%.
Aside from \validvalue, Zhou et al.~\cite{zhou2017analyzing} is able to extract specifications for type restrictions. However, we found that their heuristics that can be applied to DL document, e.g., \q{[something] be {not} [SpecClassName]}, extract at most 28 constraints (0.2\% of \tool constraints). This results in their best recall of 12.3\%, while \tool has a recall of \recalltotal\%.
Overall, \shape constraints are DL-specific that \tool extracts while existing techniques do not consider. 
The results show that \tool complements existing constraint-extraction techniques by extracting DL-specific constraints automatically.

\subsection{RQ3: Bug Detection Results}
\label{sec:bugresults}
\noindent\textbf{Approach: }
We demonstrate  the effectiveness of \tool's constraint extraction using the constraints to guide input generation to detect bugs in  API documents and library code. 
For documentation bugs, \tool detects  inconsistencies within API documents when extracting  constraints, which will be discussed later in this section. 
For library code bugs, we use all \totalconstraint (Table~\ref{tab:constraint_quality}) constraints extracted by \tool
to generate inputs for API functions that have at least one extracted constraint.
Table~\ref{tab:constraint_quality} shows the numbers of these API functions (row \textit{\#APIs with extracted constr.}).
We set \code{maxIter} to 2,000. 
For each API function, 
\tool generates $2,000$ test inputs 
(1,000 conforming 
and 1,000 violating inputs),
evaluates them, and returns bug-triggering inputs that cause serious failures (details in Section~\ref{sec:fuzzing_process}). 
We manually examine those bug-triggering inputs to check if they reveal real bugs.
For those inputs that still trigger the same failures in the nightly version, we report the bugs to the developers.

We implement an unguided input generation tool as the \emph{baseline}.
The only difference between \tool and the baseline is that the baseline has no knowledge of constraints.  
Specifically, the baseline
generates 2,000 random inputs for 
parameters 
without any constraint knowledge.  
For a fair comparison, we convert the generated array inputs to tensors assuming that the baseline
minimally knows 
which input arguments
should be tensors. 
Without this conversion, non-tensor input arguments are trivially rejected by PyTorch and MXNet, thus very ineffective in exercising the code in depth.

The extracted constraints can be used together with other input-generation tools to improve their testing effectiveness. In this paper, 
we choose to implement our own baseline instead of using existing fuzzers~\cite{afl,honggfuzz,libfuzzer} 
such as  AFL~\cite{afl} for practical reasons.  These fuzzers cannot test  Python code: the most popular language for DL. 
Moreover, these fuzzers require code coverage, which is currently unavailable across Python and C++. 
Instead of code coverage, \tool  uses constraints extracted from documents to guide the testing of both Python and C++ code, 
by generating inputs for the Python API functions, in which C++ code is invoked. 
In addition, existing fuzzers~\cite{afl,honggfuzz,libfuzzer} 
generate inputs in the format of a sequence of byte arrays. Randomly mutating some bytes 
is unlikely to generate valid DL-specific inputs.
Our baseline is similar to AFL with two enhancements: (1) knowledge of tensors and (2) automatically testing Python and C++ code.

\begin{table}[t]
\centering
\caption{
Number of \textbf{verified new / new / all } bugs (buggy APIs) 
}
\label{tab:num_bugs_summary}

\resizebox{\linewidth}{!}{
\begin{tabular}{@{}llr@{ }r@{ }l@{ }r@{ }r@{ }rr@{ }r@{ }rr@{ }r@{ }r@{}}
\toprule
\textbf{Approach} & & \multicolumn{3}{l}{\textbf{TensorFlow}} & \multicolumn{3}{c}{\textbf{PyTorch}} & \multicolumn{3}{c}{\textbf{MXNet}} & \multicolumn{3}{c}{\textbf{Total}} \\
\midrule
\textbf{Baseline} &
& 22~/& 26~/& 41~(79)                             
& 6~/& 6~/& 7~(8)            
& 6~/& 9~/& 11~(21)         
& \baseNewVeri~/& \baseNewAll~/& \baseAll~(\numAPICrashBase)
\\

\midrule
\multirow{3}{*}{\begin{tabular}{@{}l@{}}\textbf{\tool}  \end{tabular}}
& All
& 31~/& 38~/& 61~(\numAPICrashTF)                              
& 13~/& 13~/& 18~(\numAPICrashPT)           
& 10~/& ~12~/& 15~(\numAPICrashMX)         
& \oursNewVeri~/& \oursNewAll~/& \textbf{\oursAll~(\numAPICrashTotal)}  
\\

\cmidrule(ll){2-14}

& ~CI     
&  21~/& 28~/& 47~(93)                                
&  11~/&  11~/&  14~(23)                                    
&  10~/&  12~/&  14~(27)                                  
& 42~/& 51~/& \oursCIAll~(143)                          
\\

& ~VI     
& 28~/& 32~/& 51~(83)                                     
& 13~/&  13~/& 18~(25)                                    
& 8~/&  10~/&  12~(27)                                 
& 49~/& 55~/& \oursVIAll~(135)   
\\

\bottomrule
\end{tabular}
}

\vspace{-2mm}
\end{table}

\vspace{0.03in}
\noindent\textbf{Bugs in libraries code:}
Table~\ref{tab:num_bugs_summary} presents the number of \textit{verified new  bugs, new bugs,  all bugs}, 
found by the baseline 
and \tool.
A bug is verified if it has been fixed or confirmed by the developers. 
A new bug refers to a previously unknown bug that we reported. 
The unverified new bugs are reproducible and waiting for developers' responses. 
We count one bug for each required fixing location.

\tool detects \oursAll bugs, including \oursNewAll previously unknown bugs, \oursNewVeri of which have been verified by the developers (\newBugFixed fixed and \newBugConfirmed confirmed).
Of the \newBugFixed fixed bugs,
\newBugFixedC are fixed in C++,
\newBugFixedPy are fixed in Python,
\newBugFixedboth are fixed in both, and 
\newBugFixedunknown is 
fixed silently
after we reported.
The \oursAll  bugs cause \numAPICrashTotal APIs to fail because 
one bug can cause failures in multiple APIs but are fixed in one location. We count them as \oursAll instead of \numAPICrashTotal bugs. 
Of the \numAPICrashTotal buggy APIs, 
\numAPICrashDep 
have parameters with constraint dependencies.
\tool has also detected \knownBugFixed 
($\oursAll-\oursNewAll$)
known bugs that have been fixed in the nightly versions.

The baseline detects only \baseAll bugs with \baseNewAll new bugs causing \numAPICrashBase  APIs to crash.
\tool detects \oursMinusbase 
bugs that the baseline cannot while missing  \missingBugBase bugs found by the baseline due to the randomness of the input generation process.

\vspace{0.03in}
\noindent\textbf{False positives:} Only 3 out of 66
newly reported bugs receive ``won't fix'' responses from the developers.
They claim such inputs are not supported,
which is not stated in the document.
We do not count these 3 bugs in our results.

\tool uses the automatically extracted constraints without any manual examination. 
It is possible that documents themselves are incorrect or incomplete, causing incorrect constraints to be extracted,  leading to  false positives,  where the code is correct, but the API documentation is incorrect. Since we focus on severe bugs such as crashes, all detected bugs are in library code bugs, as well said by a developer after we reported a crash bug ``\emph{A segmentation fault is never OK and we should fix it with high priority.}"

\vspace{0.03in}
\noindent\textbf{Conforming and violating inputs:} 
\tool generates both conforming inputs (CIs) and violating inputs (VIs). 
Rows ``CI'' and ``VI'' in Table~\ref{tab:num_bugs_summary} present the breakdown of the bug detection for CIs and VIs
with \code{conform\_ratio = 50\%}.
We find that \tool is insensitive to conform\_ratio. When it is
between  
20\%--60\%
(with a 10\% increment): the number of detected
 bugs differs by at most one. Thus, we use conform\_ratio = 50\%
 as the default ratio to be more general.

The results show that the CIs
alone, with only 50\% (1,000) of the number of test inputs of the baseline (2,000),
finds more bugs (\oursCIAll bugs) than the baseline (\baseAll bugs), and the VIs 
alone (with 50\% of the test inputs) 
finds more bugs (\oursVIAll bugs) than the baseline.
We manually verify the generated CIs and VIs:
out of \oursCIAll CI bugs we found, $57$ of them are caused by valid inputs conforming to the ground truth constraints. The rest of the CI bugs are caused by invalid inputs generated by conforming to inaccurate constraints; out of \oursVIAll VI bugs we found, all of them are caused by invalid inputs violating the ground truth constraints.

Many bugs are detected by both CIs and VIs (comparing the ``All'' row with the ``CI'' and ``VI'' rows in Table~\ref{tab:num_bugs_summary})
because \tool violates the constraints of one parameter only when
generating VIs.
When a crash is caused by one of the conforming parameters of a VI, it is likely to be triggered
by a CI also.
However, both CIs and VIs 
trigger bugs in unique APIs,
thus both 
are effective in finding bugs.

Without the constraints, a baseline is much worse than the results from any of the ratio setups. 
Table~\ref{tab:num_bugs_summary} shows that a \emph{key contribution of our work is the ability to extract constraints from documents}. 
One cannot choose to focus on valid or invalid inputs without knowing the definition of valid inputs for an API. \tool enables this choice since it extracts input constraints automatically.

\vspace{0.03in}
\noindent\textbf{Impact of fuzzer's nondeterminism: }Since the fuzzing process is non-deterministic, we 
investigate the impact of this nondeterminism to ensure the validity of our results, i.e., the reported improvement of \tool is not due the randomness in the fuzzer. Our results suggest that it is statistically significant that \tool (which is guided by constraints) outperforms a baseline fuzzer.
Specifically, we repeat the fuzzing experiment (with the same set of extracted constraints) eight times. 
For each run, we generate 2,000 test inputs for the baseline and 2,000 test inputs for \tool for all APIs from the three libraries. In each run, we use the same random seed for both the baseline and \tool. Different runs use different seeds. 
Since it requires significant manual effort to inspect the detected bugs from all eight runs, which are 2,301 API crashes to examine, we use the number of buggy APIs (i.e., the number of APIs that crash) 
to indicate the fuzzers' effectiveness.
Overall, among the eight runs, \tool on average detects 172.0 buggy APIs 
while the baseline on average detects 115.6 buggy APIs.
We perform the Mann-Whitney U-test and confirm the improvement is statistically significant with a p-value of 0.0004 and the Cohen’s d effect size of 8.96 (effect size more than 2.0 is huge). The detailed results are in~\cite{supplementary-material}. 

\vspace{0.03in}
\noindent\textbf{Bugs in API documents:}
\tool 
detects three types of documentation bugs: (1) formatting bugs~(e.g., indentation issue); (2) signature-description mismatch~(the description refers to parameters that are not specified in the API signature); and (3) unclear constraint dependency~(Section~\ref{approach_constr_extractor}).
\tool detects \docBugAll previously unknown documentation bugs in 46 APIs~(\docBugFormat formatting bugs, \docBugInconsis signature-description mismatches, 
\docBugDep unclear constraint dependencies). 
Majority (\docBugVeri of \docBugAll) are fixed or confirmed after we report,
indicating that 
\tool detects documentation bugs that developers care to fix.

\vspace{0.03in}
\noindent\textbf{Bug examples:}
We present 
three 
bugs detected by \tool that the baseline fails to detect. All of them have been fixed by developers after we report them. \textbf{Bug 3} is also reported as a security vulnerability
CVE-2020-15265.

\noindent\textbf{Bug 1:} The previously unknown bug in TensorFlow \code{tf.nn.max\_pool3d}
discussed in the Introduction (Fig.~\ref{fig:API_example_all}).

\vspace{0.03in}
\noindent\textbf{Bug 2:}
In the API \code{tf.image.combined\_non\_max\_suppression}, \tool 
detects a previously-unknown bug and triggers a memory overflow 
by passing a large value of 311452676677046672 for the parameter \code{max\_total\_size}.
To successfully trigger this bug,
\tool needs to generate correct shaped values for parameters \code{boxes} and \code{scores}.
Specifically, the parameter \code{boxes} needs to be 4-D with the size of the last dimension equals to 4 while the parameter \code{scores} needs to be 3-D.
\tool also needs to follow the dependencies between those two parameters -- the sizes of the first two dimensions of \code{boxes} and \code{scores} need to be the same. 
\tool does this by extracting relevant \shape constraints and the dependencies correctly from the API document.
Without such knowledge, random input generation fails to produce valid input for \code{boxes} and \code{scores} to trigger this bug.

\vspace{0.03in}
\noindent\textbf{Bug 3:}
\tool triggers a segmentation fault bug in the TensorFlow API \code{tf.quantization.quantize\_and\_dequantize} using an \code{input} tensor of any shape with an out-of-bound \code{axis} value (i.e., the value of \code{axis} is larger than the number of \code{input} dimensions). After we report the bug, TensorFlow developers report this as a security vulnerability
CVE-2020-15265
to the national vulnerability database (NVD). The extracted constraints enable \tool to trigger this bug by ensuring the generation of many valid input values for all parameters of this API other than the \code{axis} values. 
For example, the extracted constraints contain the boolean type for parameters \code{narrow\_range}, \code{range\_given}, and \code{signed\_input} and the valid values (\q{\code{HALF\_TO\_EVEN}} and \q{\code{HALF\_UP}}) for parameter \code{round\_mode}, 
which help \tool generate valid values for these parameters. 
In contrast, since the baseline is unaware of these constraints, the baseline generates invalid values for these parameters, which are rejected by TensorFlow's input validation code, therefore avoiding exposing this security vulnerability.

\begin{figure}[t]
	\centering
	\includegraphics[width=0.9\linewidth]{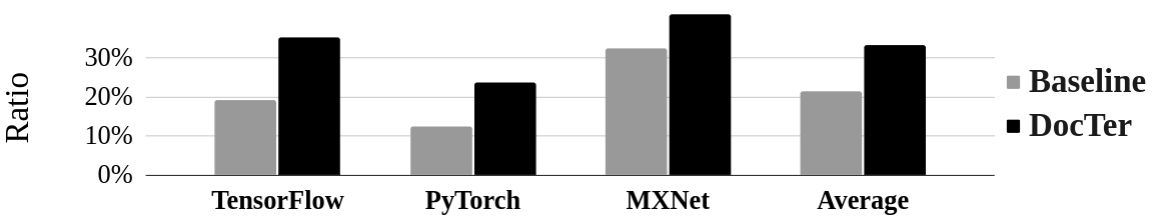}
	\caption{Ratio of passing inputs
	}
	\label{fig:valid_inputs}
	\vspace{-3mm}
\end{figure}

\subsection{RQ4: Valid-Input Generation Results}
\noindent\textbf{Approach:} As discussed in the Introduction, generating valid inputs is essential to exercise the core functionality of the API function. While \tool attempts to generate CIs, these 
CIs may still be invalid if the constraints extracted are incorrect or incomplete. We study the percentage of generated CIs
that are valid inputs. 
We compute the ratio out of
1,000 CIs (\code{confirming\_ratio = 50\%}) with the first 1,000 baseline inputs for each API function. 
Since manually examining the validity of all inputs 
is impractical and the validity checking of mature projects (e.g., our subjects) is generally reliable,
we make a reasonable
approximation by counting 
the number of passing inputs whose executions 
terminate normally.

\vspace{0.03in}
\noindent\textbf{Results:} 
Fig.~\ref{fig:valid_inputs} presents the ratio of 
passing
inputs for each subject and the average. On average, \tool achieves 33.4\% ratio of passing inputs, which outperforms the baseline (21.5\%)
by generating 55.3\% more 
passing
inputs.
The results suggest that \tool is 
more 
effective in generating valid inputs than the baseline 
to detect more bugs. 
Although \tool outperforms the baseline,
the ratio of passing inputs is still low (33.4\%),
because 
API documents are often incomplete. 
\tool might convince developers to write more complete documents since documents can help them find bugs.

\section{Threats to Validity}
\label{sec:threats}

\vspace{0.03in}
\noindent\textbf{Complex constraints:}
\tool
does not 
work with
complex constraints
that require 
a nested structure 
or indirect dependency with the constraints of another parameter.
However, 
these complex constraints 
are uncommon in DL libraries (appeared in only 6.4\% of our sampled parameters).

\vspace{0.03in}
\noindent\textbf{Testing Python and C++ code:}
Since DL libraries' core computations are in C++, it may appear to be more reasonable to directly test C++ code. However, since Python APIs are the most popular for DL, testing them is testing the common use cases. 
\tool tests  Python APIs which invoke the computations in C++, so \tool  finds bugs in both Python (\newBugFixedPytotal bugs) code and C++ (\newBugFixedCtotal bugs) code.

\vspace{0.03in}
\noindent\textbf{Manual annotations:} There is a one-time cost of up to \manhour man hours 
of manually annotating 30\% of parameters with their ACs (Section~\ref{sec:pattern_miner}). Since the \tool-generated rules are applicable to other libraries and versions (Section~\ref{extraction-results}), this one-time cost is reasonable.
Such manual annotation cost is widely accepted in other domains (e.g., supervised learning). 
Moreover, to minimize biases with the manual annotation, our process involves three co-authors. Two authors independently annotate with 98.2\% agreement. All disagreements are resolved with a third author to reach a consensus.

\balance
\vspace{-1.5mm}
\section{Related Work}
\label{sec:related}
 
\tool is the first technique to extract DL-specific constraints from API documentation, and the first DL library testing technique that is guided by such input constraints.

\vspace{0.03in}
\noindent\textbf{Constraint extraction:} 
Existing constraint-extraction techniques are insufficient for extracting DL-specific constraints~\cite{pytype, jdoctor, c2s, wong2015dase, RTFM-advance, tcomment, zhou2017analyzing}, because they miss most of the DL-specific constraints, cannot analyze across Python and C++ code, or requires formal specifications. 
Many existing techniques~\cite{jdoctor,wong2015dase,tcomment, indicator, toradocu}  use a handful of manually-designed rules to extract constraints. 
Instead, \tool 
uses subtree matching to construct rules to extract constraints automatically.

\vspace{0.03in}
\noindent\textbf{Analyzing software text to detect bugs:} 
Prior work leverages documents~\cite{RTFM-advance} and comments~\cite{icomment, acomment, zhou2017analyzing, tcomment} to detect inconsistency bugs between code and its specifications. 
Some prior work 
translates software specifications into assertions~\cite{liu2014automatic} and oracles~\cite{toradocu, swami}. 
Different from these techniques, \tool uses frequent subtree mining and association rule learning to extract constraints from API documents to guide input generation for testing DL libraries.

\vspace{0.03in}
\noindent\textbf{Testing DL libraries and fuzzing:} 
\label{sec:related_testingDLlibs} 
The constraints extracted may be used to improve existing testing techniques. 
The 
DL library testing techniques focus on addressing the test oracle challenge, by using differential testing~\cite{cradle, discrepancies, wang:fse20, MultipleTesting18,probfuzz, Audee}
or oracle approximation~\cite{test-dl-lib, 8802818}. 
\tool uses crashes instead and addresses the challenge of obtaining input constraints automatically.  

Existing techniques are designed to detect specific types of bugs such as shape-related (e.g., tensor shape mismatch)~\cite{probfuzz, shape_tf}, numerical~\cite{Audee, probfuzz} (e.g., returns \code{NaN/Inf}), decreased accuracy~\cite{probfuzz}, and performance~\cite{perf-bug-ML}. 
On the other hand, \tool finds general bugs that lead to serious crashes.

TensorFlow developers use OSS-Fuzz~\cite{oss-fuzz} along with libFuzzer~\cite{libfuzzer} to test only $19$ TensorFlow's C++ API functions.
It requires developers to manually encode 
test inputs
from the 
byte-arrays returned by libFuzzer.
This would take a prohibitive amount of manual effort to test the 
\totalapiwithconstraint APIs that \tool tests.

Fuzzers~\cite{afl,FuzzFactory,libfuzzer}
have been adopted to test non-DL  libraries~\cite{LemieuxS18,PengSP18}.  
They would not work well for DL libraries
(Section~\ref{sec:bugresults}). 
Since Randoop~\cite{randoop} works only for a statically typed language (e.g., Java), it 
 would fail to create valid dynamically typed objects for
Python
(the most popular language 
for DL~\cite{popularlanguage}).

\vspace{0.03in}
\noindent\textbf{Testing DL models:}
Many fuzzing techniques test the robustness of DL \emph{models} instead of DL \emph{libraries} \advcite. \tool tests DL libraries since testing DL models alone is insufficient as 
DL libraries contain bugs~\cite{DL_bug_char, 2020-Humbatova-ICSE, cradle}.

\section{Conclusion}

We propose 
\tool, which features a novel method to derive general rules to translate API documents to precise parameter constraints. We apply these rules to popular DL libraries to extract a large number of DL-specific constraints. 
We use the constraints to guide the input generation of DL API functions. 
The constraints enable \tool to generate valid and invalid inputs
to detect more bugs in code and documents.

\section*{Acknowledgement}
The authors thank the anonymous reviewers for their invaluable
feedback. The research is partially supported by NSF 1901242 award.


\balance
\bibliographystyle{ACM-Reference-Format}
\bibliography{paper} 


\begin{thebibliography}{62}


\ifx \showCODEN    \undefined \def \showCODEN     #1{\unskip}     \fi
\ifx \showDOI      \undefined \def \showDOI       #1{#1}\fi
\ifx \showISBNx    \undefined \def \showISBNx     #1{\unskip}     \fi
\ifx \showISBNxiii \undefined \def \showISBNxiii  #1{\unskip}     \fi
\ifx \showISSN     \undefined \def \showISSN      #1{\unskip}     \fi
\ifx \showLCCN     \undefined \def \showLCCN      #1{\unskip}     \fi
\ifx \shownote     \undefined \def \shownote      #1{#1}          \fi
\ifx \showarticletitle \undefined \def \showarticletitle #1{#1}   \fi
\ifx \showURL      \undefined \def \showURL       {\relax}        \fi
\providecommand\bibfield[2]{#2}
\providecommand\bibinfo[2]{#2}
\providecommand\natexlab[1]{#1}
\providecommand\showeprint[2][]{arXiv:#2}

\bibitem[\protect\citeauthoryear{??}{jml}{1999}]%
        {jml}
 \bibinfo{year}{1999}\natexlab{}.
\newblock \bibinfo{booktitle}{\emph{The Java Modeling Language (JML)}}.
\newblock
\urldef\tempurl%
\url{"https://www.cs.ucf.edu/~leavens/JML/examples.shtml"}
\showURL{%
\tempurl}


\bibitem[\protect\citeauthoryear{??}{bea}{2004}]%
        {beautifulsoup}
 \bibinfo{year}{2004}\natexlab{}.
\newblock \bibinfo{booktitle}{\emph{Beautiful Soup}}.
\newblock
\urldef\tempurl%
\url{https://www.crummy.com/software/BeautifulSoup/bs4/doc/}
\showURL{%
\tempurl}


\bibitem[\protect\citeauthoryear{??}{afl}{2013}]%
        {afl}
 \bibinfo{year}{2013}\natexlab{}.
\newblock \bibinfo{title}{American Fuzzy Lop}.
\newblock
\newblock
\urldef\tempurl%
\url{http://lcamtuf.coredump.cx/afl/}
\showURL{%
\tempurl}


\bibitem[\protect\citeauthoryear{??}{dep}{2014}]%
        {dep-standard}
 \bibinfo{year}{2014}\natexlab{}.
\newblock \bibinfo{title}{Universal Dependencies}.
\newblock
\newblock
\urldef\tempurl%
\url{https://universaldependencies.org/}
\showURL{%
\tempurl}


\bibitem[\protect\citeauthoryear{??}{lib}{2015}]%
        {libfuzzer}
 \bibinfo{year}{2015}\natexlab{}.
\newblock \bibinfo{title}{libFuzzer – a library for coverage-guided fuzz
  testing.}
\newblock
\newblock
\urldef\tempurl%
\url{http://llvm.org/docs/LibFuzzer.html}
\showURL{%
\tempurl}


\bibitem[\protect\citeauthoryear{??}{oss}{2016}]%
        {oss-fuzz}
 \bibinfo{year}{2016}\natexlab{}.
\newblock \bibinfo{booktitle}{\emph{OSS-Fuzz}}.
\newblock
\urldef\tempurl%
\url{https://github.com/google/oss-fuzz}
\showURL{%
\tempurl}


\bibitem[\protect\citeauthoryear{??}{pyt}{2016}]%
        {pytype}
 \bibinfo{year}{2016}\natexlab{}.
\newblock \bibinfo{booktitle}{\emph{pytype}}.
\newblock
\urldef\tempurl%
\url{"https://github.com/google/pytype"}
\showURL{%
\tempurl}


\bibitem[\protect\citeauthoryear{??}{pop}{2017}]%
        {popularlanguage}
 \bibinfo{year}{2017}\natexlab{}.
\newblock \bibinfo{title}{What is the best programming language for Machine
  Learning?}
\newblock
  \bibinfo{howpublished}{\url{https://towardsdatascience.com/what-is-the-best-programming-language-for-machine-learning-a745c156d6b7}}.
\newblock


\bibitem[\protect\citeauthoryear{??}{Fuz}{2019}]%
        {FuzzFactory}
 \bibinfo{year}{2019}\natexlab{}.
\newblock \bibinfo{booktitle}{\emph{FuzzFactory: Domain-Specific Fuzzing with
  Waypoints}}.
\newblock
\urldef\tempurl%
\url{"https://github.com/rohanpadhye/fuzzfactory"}
\showURL{%
\tempurl}


\bibitem[\protect\citeauthoryear{??}{mx_}{2019}]%
        {mx_dtype}
 \bibinfo{year}{2019}\natexlab{}.
\newblock \bibinfo{title}{incubator-mxnet}.
\newblock
\newblock
\urldef\tempurl%
\url{https://github.com/apache/incubator-mxnet/blob/1.6.0/python/mxnet/ndarray/ndarray.py#L64-L74}
\showURL{%
\tempurl}


\bibitem[\protect\citeauthoryear{??}{pt_}{2019}]%
        {pt_dtype}
 \bibinfo{year}{2019}\natexlab{}.
\newblock \bibinfo{title}{torch.Tensor}.
\newblock
\newblock
\urldef\tempurl%
\url{https://pytorch.org/docs/1.5.0/tensors.html}
\showURL{%
\tempurl}


\bibitem[\protect\citeauthoryear{??}{tf_}{2020}]%
        {tf_dtype}
 \bibinfo{year}{2020}\natexlab{}.
\newblock \bibinfo{title}{tf.dtypes.DType}.
\newblock
\newblock
\urldef\tempurl%
\url{https://www.tensorflow.org/versions/r2.1/api_docs/python/tf/dtypes/DType}
\showURL{%
\tempurl}


\bibitem[\protect\citeauthoryear{??}{sup}{2022}]%
        {supplementary-material}
 \bibinfo{year}{2022}\natexlab{}.
\newblock \bibinfo{title}{\tool's Supplementary Material}.
\newblock
\newblock
\urldef\tempurl%
\url{https://github.com/lin-tan/DocTer}
\showURL{%
\tempurl}


\bibitem[\protect\citeauthoryear{Abadi, Barham, Chen, Chen, Davis, Dean, Devin,
  Ghemawat, Irving, Isard, et~al\mbox{.}}{Abadi et~al\mbox{.}}{2016}]%
        {abadi2016tensorflow}
\bibfield{author}{\bibinfo{person}{Mart{\'\i}n Abadi}, \bibinfo{person}{Paul
  Barham}, \bibinfo{person}{Jianmin Chen}, \bibinfo{person}{Zhifeng Chen},
  \bibinfo{person}{Andy Davis}, \bibinfo{person}{Jeffrey Dean},
  \bibinfo{person}{Matthieu Devin}, \bibinfo{person}{Sanjay Ghemawat},
  \bibinfo{person}{Geoffrey Irving}, \bibinfo{person}{Michael Isard},
  {et~al\mbox{.}}} \bibinfo{year}{2016}\natexlab{}.
\newblock \showarticletitle{Tensorflow: A system for large-scale machine
  learning}. In \bibinfo{booktitle}{\emph{12th $\{$USENIX$\}$ symposium on
  operating systems design and implementation ($\{$OSDI$\}$ 16)}}.
  \bibinfo{pages}{265--283}.
\newblock


\bibitem[\protect\citeauthoryear{Bird, Loper, and Klein}{Bird
  et~al\mbox{.}}{2009}]%
        {nltk}
\bibfield{author}{\bibinfo{person}{Steven Bird}, \bibinfo{person}{Edward
  Loper}, {and} \bibinfo{person}{Ewan Klein}.} \bibinfo{year}{2009}\natexlab{}.
\newblock \bibinfo{booktitle}{\emph{Natural Language Processing with Python}}.
\newblock \bibinfo{publisher}{O'Reilly Media Inc}.
\newblock


\bibitem[\protect\citeauthoryear{Blasi, Goffi, Kuznetsov, Gorla, Ernst,
  Pezz{\`e}, and Castellanos}{Blasi et~al\mbox{.}}{2018}]%
        {jdoctor}
\bibfield{author}{\bibinfo{person}{Arianna Blasi}, \bibinfo{person}{Alberto
  Goffi}, \bibinfo{person}{Konstantin Kuznetsov}, \bibinfo{person}{Alessandra
  Gorla}, \bibinfo{person}{Michael~D Ernst}, \bibinfo{person}{Mauro Pezz{\`e}},
  {and} \bibinfo{person}{Sergio~Delgado Castellanos}.}
  \bibinfo{year}{2018}\natexlab{}.
\newblock \showarticletitle{Translating code comments to procedure
  specifications}. In \bibinfo{booktitle}{\emph{Proceedings of the 27th ACM
  SIGSOFT International Symposium on Software Testing and Analysis}}.
  \bibinfo{pages}{242--253}.
\newblock


\bibitem[\protect\citeauthoryear{Chen, Li, Li, Lin, Wang, Wang, Xiao, Xu,
  Zhang, and Zhang}{Chen et~al\mbox{.}}{2015}]%
        {chen2015mxnet}
\bibfield{author}{\bibinfo{person}{Tianqi Chen}, \bibinfo{person}{Mu Li},
  \bibinfo{person}{Yutian Li}, \bibinfo{person}{Min Lin},
  \bibinfo{person}{Naiyan Wang}, \bibinfo{person}{Minjie Wang},
  \bibinfo{person}{Tianjun Xiao}, \bibinfo{person}{Bing Xu},
  \bibinfo{person}{Chiyuan Zhang}, {and} \bibinfo{person}{Zheng Zhang}.}
  \bibinfo{year}{2015}\natexlab{}.
\newblock \bibinfo{title}{MXNet: A Flexible and Efficient Machine Learning
  Library for Heterogeneous Distributed Systems}.
\newblock
\newblock
\showeprint[arxiv]{1512.01274}~[cs.DC]


\bibitem[\protect\citeauthoryear{Dutta, Legunsen, Huang, and Misailovic}{Dutta
  et~al\mbox{.}}{2018}]%
        {probfuzz}
\bibfield{author}{\bibinfo{person}{Saikat Dutta}, \bibinfo{person}{Owolabi
  Legunsen}, \bibinfo{person}{Zixin Huang}, {and} \bibinfo{person}{Sasa
  Misailovic}.} \bibinfo{year}{2018}\natexlab{}.
\newblock \showarticletitle{Testing Probabilistic Programming Systems.}
  \emph{(\bibinfo{series}{ESEC/FSE 2018})}. \bibinfo{publisher}{Association for
  Computing Machinery}, \bibinfo{address}{New York, NY, USA}.
\newblock
\showISBNx{9781450355735}
\urldef\tempurl%
\url{https://doi.org/10.1145/3236024.3236057}
\showDOI{\tempurl}


\bibitem[\protect\citeauthoryear{Fraser and Arcuri}{Fraser and Arcuri}{2013}]%
        {TSE12_EvoSuite}
\bibfield{author}{\bibinfo{person}{Gordon Fraser} {and} \bibinfo{person}{Andrea
  Arcuri}.} \bibinfo{year}{2013}\natexlab{}.
\newblock \showarticletitle{Whole Test Suite Generation}.
\newblock \bibinfo{journal}{\emph{IEEE Transactions on Software Engineering}}
  \bibinfo{volume}{39}, \bibinfo{number}{2} (\bibinfo{date}{feb.}
  \bibinfo{year}{2013}), \bibinfo{pages}{276 --291}.
\newblock


\bibitem[\protect\citeauthoryear{Gao, Saha, Prasad, and Roychoudhury}{Gao
  et~al\mbox{.}}{2020}]%
        {2020FuzzTB}
\bibfield{author}{\bibinfo{person}{Xiang Gao}, \bibinfo{person}{Ripon~K Saha},
  \bibinfo{person}{Mukul~R Prasad}, {and} \bibinfo{person}{Abhik
  Roychoudhury}.} \bibinfo{year}{2020}\natexlab{}.
\newblock \showarticletitle{Fuzz Testing based Data Augmentation to Improve
  Robustness of Deep Neural Networks}. In \bibinfo{booktitle}{\emph{Proceedings
  of the 42nd International Conference on Software Engineering (ICSE '20)}}.
\newblock


\bibitem[\protect\citeauthoryear{Godefroid, Kiezun, and Levin}{Godefroid
  et~al\mbox{.}}{2008}]%
        {grammarFuzzing}
\bibfield{author}{\bibinfo{person}{P. Godefroid}, \bibinfo{person}{A. Kiezun},
  {and} \bibinfo{person}{M.~Y. Levin}.} \bibinfo{year}{2008}\natexlab{}.
\newblock \showarticletitle{Grammar-based Whitebox Fuzzing}. In
  \bibinfo{booktitle}{\emph{Proceedings of the ACM SIGPLAN conference on
  Programming language design and implementation}}. \bibinfo{pages}{206--215}.
\newblock


\bibitem[\protect\citeauthoryear{Goffi, Gorla, Ernst, and Pezz{\`e}}{Goffi
  et~al\mbox{.}}{2016}]%
        {toradocu}
\bibfield{author}{\bibinfo{person}{Alberto Goffi}, \bibinfo{person}{Alessandra
  Gorla}, \bibinfo{person}{Michael~D Ernst}, {and} \bibinfo{person}{Mauro
  Pezz{\`e}}.} \bibinfo{year}{2016}\natexlab{}.
\newblock \showarticletitle{Automatic generation of oracles for exceptional
  behaviors}. In \bibinfo{booktitle}{\emph{Proceedings of the 25th
  international symposium on software testing and analysis}}.
  \bibinfo{pages}{213--224}.
\newblock


\bibitem[\protect\citeauthoryear{Guo, Jiang, Zhao, Chen, and Sun}{Guo
  et~al\mbox{.}}{2018}]%
        {dlfuzz}
\bibfield{author}{\bibinfo{person}{Jianmin Guo}, \bibinfo{person}{Yu Jiang},
  \bibinfo{person}{Yue Zhao}, \bibinfo{person}{Quan Chen}, {and}
  \bibinfo{person}{Jiaguang Sun}.} \bibinfo{year}{2018}\natexlab{}.
\newblock \showarticletitle{DLFuzz: Differential Fuzzing Testing of Deep
  Learning Systems}. In \bibinfo{booktitle}{\emph{Proceedings of the 2018 26th
  ACM Joint Meeting on European Software Engineering Conference and Symposium
  on the Foundations of Software Engineering}} (Lake Buena Vista, FL, USA)
  \emph{(\bibinfo{series}{ESEC/FSE 2018})}. \bibinfo{publisher}{ACM},
  \bibinfo{address}{New York, NY, USA}, \bibinfo{pages}{739--743}.
\newblock
\showISBNx{978-1-4503-5573-5}
\urldef\tempurl%
\url{https://doi.org/10.1145/3236024.3264835}
\showDOI{\tempurl}


\bibitem[\protect\citeauthoryear{Guo, Xie, Li, Zhang, Liu, Li, and Shen}{Guo
  et~al\mbox{.}}{2020}]%
        {Audee}
\bibfield{author}{\bibinfo{person}{Qianyu Guo}, \bibinfo{person}{Xiaofei Xie},
  \bibinfo{person}{Yi Li}, \bibinfo{person}{Xiaoyu Zhang},
  \bibinfo{person}{Yang Liu}, \bibinfo{person}{Xiaohong Li}, {and}
  \bibinfo{person}{Chao Shen}.} \bibinfo{year}{2020}\natexlab{}.
\newblock \showarticletitle{Audee: Automated Testing for Deep Learning
  Frameworks}. In \bibinfo{booktitle}{\emph{Proceedings of the 35th IEEE/ACM
  International Conference on Automated Software Engineering (ASE)}}.
\newblock


\bibitem[\protect\citeauthoryear{{Hu}, {Ma}, {Xie}, {Yu}, {Liu}, and
  {Zhao}}{{Hu} et~al\mbox{.}}{2019}]%
        {deepmutationpp}
\bibfield{author}{\bibinfo{person}{Q. {Hu}}, \bibinfo{person}{L. {Ma}},
  \bibinfo{person}{X. {Xie}}, \bibinfo{person}{B. {Yu}}, \bibinfo{person}{Y.
  {Liu}}, {and} \bibinfo{person}{J. {Zhao}}.} \bibinfo{year}{2019}\natexlab{}.
\newblock \showarticletitle{DeepMutation++: A Mutation Testing Framework for
  Deep Learning Systems}. In \bibinfo{booktitle}{\emph{2019 34th IEEE/ACM
  International Conference on Automated Software Engineering (ASE)}}.
  \bibinfo{pages}{1158--1161}.
\newblock


\bibitem[\protect\citeauthoryear{Humbatova, Jahangirova, Bavota, Riccio,
  Stocco, and Tonella}{Humbatova et~al\mbox{.}}{2020}]%
        {2020-Humbatova-ICSE}
\bibfield{author}{\bibinfo{person}{Nargiz Humbatova}, \bibinfo{person}{Gunel
  Jahangirova}, \bibinfo{person}{Gabriele Bavota}, \bibinfo{person}{Vincenzo
  Riccio}, \bibinfo{person}{Andrea Stocco}, {and} \bibinfo{person}{Paolo
  Tonella}.} \bibinfo{year}{2020}\natexlab{}.
\newblock \showarticletitle{Taxonomy of Real Faults in Deep Learning Systems}.
  In \bibinfo{booktitle}{\emph{Proceedings of 42nd International Conference on
  Software Engineering}} \emph{(\bibinfo{series}{ICSE '20})}.
  \bibinfo{publisher}{ACM}.
\newblock


\bibitem[\protect\citeauthoryear{Islam, Nguyen, Pan, and Rajan}{Islam
  et~al\mbox{.}}{2019}]%
        {DL_bug_char}
\bibfield{author}{\bibinfo{person}{Md~Johirul Islam}, \bibinfo{person}{Giang
  Nguyen}, \bibinfo{person}{Rangeet Pan}, {and} \bibinfo{person}{Hridesh
  Rajan}.} \bibinfo{year}{2019}\natexlab{}.
\newblock \showarticletitle{A Comprehensive Study on Deep Learning Bug
  Characteristics}. In \bibinfo{booktitle}{\emph{Proceedings of the 2019 27th
  ACM Joint Meeting on European Software Engineering Conference and Symposium
  on the Foundations of Software Engineering}} (Tallinn, Estonia)
  \emph{(\bibinfo{series}{ESEC/FSE 2019})}. \bibinfo{publisher}{Association for
  Computing Machinery}, \bibinfo{address}{New York, NY, USA},
  \bibinfo{pages}{510–520}.
\newblock
\showISBNx{9781450355728}
\urldef\tempurl%
\url{https://doi.org/10.1145/3338906.3338955}
\showDOI{\tempurl}


\bibitem[\protect\citeauthoryear{Lagouvardos, Dolby, Grech, Antoniadis, and
  Smaragdakis}{Lagouvardos et~al\mbox{.}}{2020}]%
        {shape_tf}
\bibfield{author}{\bibinfo{person}{Sifis Lagouvardos}, \bibinfo{person}{Julian
  Dolby}, \bibinfo{person}{Neville Grech}, \bibinfo{person}{Anastasios
  Antoniadis}, {and} \bibinfo{person}{Yannis Smaragdakis}.}
  \bibinfo{year}{2020}\natexlab{}.
\newblock \showarticletitle{Static Analysis of Shape in TensorFlow Programs.}.
  In \bibinfo{booktitle}{\emph{ECOOP 2020}}.
\newblock


\bibitem[\protect\citeauthoryear{Lemieux and Sen}{Lemieux and Sen}{2018}]%
        {LemieuxS18}
\bibfield{author}{\bibinfo{person}{Caroline Lemieux} {and}
  \bibinfo{person}{Koushik Sen}.} \bibinfo{year}{2018}\natexlab{}.
\newblock \showarticletitle{FairFuzz: a targeted mutation strategy for
  increasing greybox fuzz testing coverage.}. In
  \bibinfo{booktitle}{\emph{ASE}}, \bibfield{editor}{\bibinfo{person}{Marianne
  Huchard}, \bibinfo{person}{Christian Kästner}, {and} \bibinfo{person}{Gordon
  Fraser}} (Eds.). \bibinfo{publisher}{ACM}, \bibinfo{pages}{475--485}.
\newblock


\bibitem[\protect\citeauthoryear{Liu, Sun, Liu, Zhang, Wadhwa, Dong, and
  Wang}{Liu et~al\mbox{.}}{2014}]%
        {liu2014automatic}
\bibfield{author}{\bibinfo{person}{Shuang Liu}, \bibinfo{person}{Jun Sun},
  \bibinfo{person}{Yang Liu}, \bibinfo{person}{Yue Zhang},
  \bibinfo{person}{Bimlesh Wadhwa}, \bibinfo{person}{Jin~Song Dong}, {and}
  \bibinfo{person}{Xinyu Wang}.} \bibinfo{year}{2014}\natexlab{}.
\newblock \showarticletitle{Automatic early defects detection in use case
  documents}. In \bibinfo{booktitle}{\emph{Proceedings of the 29th ACM/IEEE
  international conference on Automated software engineering}}.
  \bibinfo{pages}{785--790}.
\newblock


\bibitem[\protect\citeauthoryear{Lv, Li, Yang, Chen, Liao, Wang, Hu, and
  Xing}{Lv et~al\mbox{.}}{2020}]%
        {RTFM-advance}
\bibfield{author}{\bibinfo{person}{Tao Lv}, \bibinfo{person}{Ruishi Li},
  \bibinfo{person}{Yi Yang}, \bibinfo{person}{Kai Chen},
  \bibinfo{person}{Xiaojing Liao}, \bibinfo{person}{XiaoFeng Wang},
  \bibinfo{person}{Peiwei Hu}, {and} \bibinfo{person}{Luyi Xing}.}
  \bibinfo{year}{2020}\natexlab{}.
\newblock \showarticletitle{RTFM! Automatic Assumption Discovery and
  Verification Derivation from Library Document for API Misuse Detection}. In
  \bibinfo{booktitle}{\emph{Proceedings of the 2020 ACM SIGSAC Conference on
  Computer and Communications Security}}. \bibinfo{pages}{1837--1852}.
\newblock


\bibitem[\protect\citeauthoryear{Majumda and Xu}{Majumda and Xu}{2007}]%
        {CESE}
\bibfield{author}{\bibinfo{person}{R. Majumda} {and} \bibinfo{person}{R. Xu}.}
  \bibinfo{year}{2007}\natexlab{}.
\newblock \showarticletitle{Directed Test Generation Using Symbolic Grammars}.
  In \bibinfo{booktitle}{\emph{Proceedings of the 22nd IEEE/ACM International
  Conference on Automated Software Engineering}}. \bibinfo{pages}{134--143}.
\newblock


\bibitem[\protect\citeauthoryear{Manning, Surdeanu, Bauer, Finkel, Bethard, and
  McClosky}{Manning et~al\mbox{.}}{2014}]%
        {corenlp}
\bibfield{author}{\bibinfo{person}{Christopher~D Manning},
  \bibinfo{person}{Mihai Surdeanu}, \bibinfo{person}{John Bauer},
  \bibinfo{person}{Jenny~Rose Finkel}, \bibinfo{person}{Steven Bethard}, {and}
  \bibinfo{person}{David McClosky}.} \bibinfo{year}{2014}\natexlab{}.
\newblock \showarticletitle{The Stanford CoreNLP natural language processing
  toolkit}. In \bibinfo{booktitle}{\emph{Proceedings of 52nd annual meeting of
  the association for computational linguistics: system demonstrations}}.
  \bibinfo{pages}{55--60}.
\newblock


\bibitem[\protect\citeauthoryear{Motwani and Brun}{Motwani and Brun}{2019}]%
        {swami}
\bibfield{author}{\bibinfo{person}{Manish Motwani} {and} \bibinfo{person}{Yuriy
  Brun}.} \bibinfo{year}{2019}\natexlab{}.
\newblock \showarticletitle{Automatically generating precise Oracles from
  structured natural language specifications}. In
  \bibinfo{booktitle}{\emph{2019 IEEE/ACM 41st International Conference on
  Software Engineering (ICSE)}}. IEEE, \bibinfo{pages}{188--199}.
\newblock


\bibitem[\protect\citeauthoryear{{Nejadgholi} and {Yang}}{{Nejadgholi} and
  {Yang}}{2019}]%
        {test-dl-lib}
\bibfield{author}{\bibinfo{person}{M. {Nejadgholi}} {and} \bibinfo{person}{J.
  {Yang}}.} \bibinfo{year}{2019}\natexlab{}.
\newblock \showarticletitle{A Study of Oracle Approximations in Testing Deep
  Learning Libraries}. In \bibinfo{booktitle}{\emph{2019 34th IEEE/ACM
  International Conference on Automated Software Engineering (ASE)}}.
  \bibinfo{pages}{785--796}.
\newblock


\bibitem[\protect\citeauthoryear{Odena, Olsson, Andersen, and Goodfellow}{Odena
  et~al\mbox{.}}{2019}]%
        {tensorfuzz}
\bibfield{author}{\bibinfo{person}{Augustus Odena}, \bibinfo{person}{Catherine
  Olsson}, \bibinfo{person}{David Andersen}, {and} \bibinfo{person}{Ian
  Goodfellow}.} \bibinfo{year}{2019}\natexlab{}.
\newblock \showarticletitle{{T}ensor{F}uzz: Debugging Neural Networks with
  Coverage-Guided Fuzzing}. In \bibinfo{booktitle}{\emph{Proceedings of the
  36th International Conference on Machine Learning}}
  \emph{(\bibinfo{series}{Proceedings of Machine Learning Research},
  Vol.~\bibinfo{volume}{97})}, \bibfield{editor}{\bibinfo{person}{Kamalika
  Chaudhuri} {and} \bibinfo{person}{Ruslan Salakhutdinov}} (Eds.).
  \bibinfo{publisher}{PMLR}, \bibinfo{address}{Long Beach, California, USA},
  \bibinfo{pages}{4901--4911}.
\newblock


\bibitem[\protect\citeauthoryear{Pacheco and Ernst}{Pacheco and Ernst}{2007}]%
        {randoop}
\bibfield{author}{\bibinfo{person}{Carlos Pacheco} {and}
  \bibinfo{person}{Michael~D. Ernst}.} \bibinfo{year}{2007}\natexlab{}.
\newblock \showarticletitle{Randoop: Feedback-directed Random Testing for
  Java}. In \bibinfo{booktitle}{\emph{Companion to the 22Nd ACM SIGPLAN
  Conference on Object-oriented Programming Systems and Applications
  Companion}} (Montreal, Quebec, Canada) \emph{(\bibinfo{series}{OOPSLA '07})}.
  \bibinfo{publisher}{ACM}, \bibinfo{address}{New York, NY, USA},
  \bibinfo{pages}{815--816}.
\newblock
\showISBNx{978-1-59593-865-7}
\urldef\tempurl%
\url{https://doi.org/10.1145/1297846.1297902}
\showDOI{\tempurl}


\bibitem[\protect\citeauthoryear{Pandita, Xiao, Zhong, Xie, Oney, and
  Paradkar}{Pandita et~al\mbox{.}}{2012}]%
        {alics}
\bibfield{author}{\bibinfo{person}{Rahul Pandita}, \bibinfo{person}{Xusheng
  Xiao}, \bibinfo{person}{Hao Zhong}, \bibinfo{person}{Tao Xie},
  \bibinfo{person}{Stephen Oney}, {and} \bibinfo{person}{Amit Paradkar}.}
  \bibinfo{year}{2012}\natexlab{}.
\newblock \showarticletitle{Inferring Method Specifications from Natural
  Language API Descriptions}. In \bibinfo{booktitle}{\emph{Proceedings of the
  34th International Conference on Software Engineering}} (Zurich, Switzerland)
  \emph{(\bibinfo{series}{ICSE '12})}. \bibinfo{publisher}{IEEE Press},
  \bibinfo{address}{Piscataway, NJ, USA}, \bibinfo{pages}{815--825}.
\newblock
\showISBNx{978-1-4673-1067-3}


\bibitem[\protect\citeauthoryear{Paszke, Gross, Massa, Lerer, Bradbury, Chanan,
  Killeen, Lin, Gimelshein, Antiga, Desmaison, Kopf, Yang, DeVito, Raison,
  Tejani, Chilamkurthy, Steiner, Fang, Bai, and Chintala}{Paszke
  et~al\mbox{.}}{2019}]%
        {NEURIPS2019_9015}
\bibfield{author}{\bibinfo{person}{Adam Paszke}, \bibinfo{person}{Sam Gross},
  \bibinfo{person}{Francisco Massa}, \bibinfo{person}{Adam Lerer},
  \bibinfo{person}{James Bradbury}, \bibinfo{person}{Gregory Chanan},
  \bibinfo{person}{Trevor Killeen}, \bibinfo{person}{Zeming Lin},
  \bibinfo{person}{Natalia Gimelshein}, \bibinfo{person}{Luca Antiga},
  \bibinfo{person}{Alban Desmaison}, \bibinfo{person}{Andreas Kopf},
  \bibinfo{person}{Edward Yang}, \bibinfo{person}{Zachary DeVito},
  \bibinfo{person}{Martin Raison}, \bibinfo{person}{Alykhan Tejani},
  \bibinfo{person}{Sasank Chilamkurthy}, \bibinfo{person}{Benoit Steiner},
  \bibinfo{person}{Lu Fang}, \bibinfo{person}{Junjie Bai}, {and}
  \bibinfo{person}{Soumith Chintala}.} \bibinfo{year}{2019}\natexlab{}.
\newblock \showarticletitle{PyTorch: An Imperative Style, High-Performance Deep
  Learning Library}.
\newblock In \bibinfo{booktitle}{\emph{Advances in Neural Information
  Processing Systems 32}}, \bibfield{editor}{\bibinfo{person}{H.~Wallach},
  \bibinfo{person}{H.~Larochelle}, \bibinfo{person}{A.~Beygelzimer},
  \bibinfo{person}{F.~d\textquotesingle Alch\'{e}-Buc},
  \bibinfo{person}{E.~Fox}, {and} \bibinfo{person}{R.~Garnett}} (Eds.).
  \bibinfo{publisher}{Curran Associates, Inc.}, \bibinfo{pages}{8024--8035}.
\newblock
\urldef\tempurl%
\url{http://papers.neurips.cc/paper/9015-pytorch-an-imperative-style-high-performance-deep-learning-library.pdf}
\showURL{%
\tempurl}


\bibitem[\protect\citeauthoryear{Pedregosa, Varoquaux, Gramfort, Michel,
  Thirion, Grisel, Blondel, Prettenhofer, Weiss, Dubourg, Vanderplas, Passos,
  Cournapeau, Brucher, Perrot, and Duchesnay}{Pedregosa et~al\mbox{.}}{2011}]%
        {scikit-learn}
\bibfield{author}{\bibinfo{person}{F. Pedregosa}, \bibinfo{person}{G.
  Varoquaux}, \bibinfo{person}{A. Gramfort}, \bibinfo{person}{V. Michel},
  \bibinfo{person}{B. Thirion}, \bibinfo{person}{O. Grisel},
  \bibinfo{person}{M. Blondel}, \bibinfo{person}{P. Prettenhofer},
  \bibinfo{person}{R. Weiss}, \bibinfo{person}{V. Dubourg}, \bibinfo{person}{J.
  Vanderplas}, \bibinfo{person}{A. Passos}, \bibinfo{person}{D. Cournapeau},
  \bibinfo{person}{M. Brucher}, \bibinfo{person}{M. Perrot}, {and}
  \bibinfo{person}{E. Duchesnay}.} \bibinfo{year}{2011}\natexlab{}.
\newblock \showarticletitle{Scikit-learn: Machine Learning in {P}ython}.
\newblock \bibinfo{journal}{\emph{Journal of Machine Learning Research}}
  \bibinfo{volume}{12} (\bibinfo{year}{2011}), \bibinfo{pages}{2825--2830}.
\newblock


\bibitem[\protect\citeauthoryear{Peng, Shoshitaishvili, and Payer}{Peng
  et~al\mbox{.}}{2018}]%
        {PengSP18}
\bibfield{author}{\bibinfo{person}{Hui Peng}, \bibinfo{person}{Yan
  Shoshitaishvili}, {and} \bibinfo{person}{Mathias Payer}.}
  \bibinfo{year}{2018}\natexlab{}.
\newblock \showarticletitle{T-Fuzz: Fuzzing by Program Transformation.}. In
  \bibinfo{booktitle}{\emph{IEEE Symposium on Security and Privacy}}.
  \bibinfo{publisher}{IEEE Computer Society}, \bibinfo{pages}{697--710}.
\newblock
\showISBNx{978-1-5386-4353-2}


\bibitem[\protect\citeauthoryear{Pham, Lutellier, Qi, and Tan}{Pham
  et~al\mbox{.}}{2019}]%
        {cradle}
\bibfield{author}{\bibinfo{person}{Hung~Viet Pham}, \bibinfo{person}{Thibaud
  Lutellier}, \bibinfo{person}{Weizhen Qi}, {and} \bibinfo{person}{Lin Tan}.}
  \bibinfo{year}{2019}\natexlab{}.
\newblock \showarticletitle{CRADLE: Cross-backend Validation to Detect and
  Localize Bugs in Deep Learning Libraries}. In
  \bibinfo{booktitle}{\emph{Proceedings of the 41st International Conference on
  Software Engineering}} (Montreal, Quebec, Canada)
  \emph{(\bibinfo{series}{ICSE '19})}. \bibinfo{publisher}{IEEE Press},
  \bibinfo{address}{Piscataway, NJ, USA}, \bibinfo{pages}{1027--1038}.
\newblock
\urldef\tempurl%
\url{https://doi.org/10.1109/ICSE.2019.00107}
\showDOI{\tempurl}


\bibitem[\protect\citeauthoryear{Srisakaokul, Wu, Astorga, Alebiosu, and
  Xie}{Srisakaokul et~al\mbox{.}}{2018}]%
        {MultipleTesting18}
\bibfield{author}{\bibinfo{person}{Siwakorn Srisakaokul},
  \bibinfo{person}{Zhengkai Wu}, \bibinfo{person}{Angello Astorga},
  \bibinfo{person}{Oreoluwa Alebiosu}, {and} \bibinfo{person}{Tao Xie}.}
  \bibinfo{year}{2018}\natexlab{}.
\newblock \showarticletitle{Multiple-Implementation Testing of Supervised
  Learning Software}. In \bibinfo{booktitle}{\emph{Proc. AAAI-18 Workshop on
  Engineering Dependable and Secure Machine Learning Systems (EDSMLS)}}.
\newblock


\bibitem[\protect\citeauthoryear{Swiecki}{Swiecki}{2015}]%
        {honggfuzz}
\bibfield{author}{\bibinfo{person}{Robert Swiecki}.}
  \bibinfo{year}{2015}\natexlab{}.
\newblock \showarticletitle{Honggfuzz: A general-purpose, easy-to-use fuzzer
  with interesting analysis options}.
\newblock \bibinfo{journal}{\emph{URl: https://github. com/google/honggfuzz}}
  (\bibinfo{year}{2015}).
\newblock


\bibitem[\protect\citeauthoryear{Tan, Yuan, Krishna, and Zhou}{Tan
  et~al\mbox{.}}{2007}]%
        {icomment}
\bibfield{author}{\bibinfo{person}{Lin Tan}, \bibinfo{person}{Ding Yuan},
  \bibinfo{person}{Gopal Krishna}, {and} \bibinfo{person}{Yuanyuan Zhou}.}
  \bibinfo{year}{2007}\natexlab{}.
\newblock \showarticletitle{/*Icomment: Bugs or Bad Comments?*/}. In
  \bibinfo{booktitle}{\emph{Proceedings of Twenty-first ACM SIGOPS Symposium on
  Operating Systems Principles}} (Stevenson, Washington, USA)
  \emph{(\bibinfo{series}{SOSP '07})}. \bibinfo{publisher}{ACM},
  \bibinfo{address}{New York, NY, USA}, \bibinfo{pages}{145--158}.
\newblock
\showISBNx{978-1-59593-591-5}
\urldef\tempurl%
\url{https://doi.org/10.1145/1294261.1294276}
\showDOI{\tempurl}


\bibitem[\protect\citeauthoryear{Tan, Zhou, and Padioleau}{Tan
  et~al\mbox{.}}{2011}]%
        {acomment}
\bibfield{author}{\bibinfo{person}{Lin Tan}, \bibinfo{person}{Yuanyuan Zhou},
  {and} \bibinfo{person}{Yoann Padioleau}.} \bibinfo{year}{2011}\natexlab{}.
\newblock \showarticletitle{aComment: Mining Annotations from Comments and Code
  to Detect Interrupt Related Concurrency Bugs}. In
  \bibinfo{booktitle}{\emph{Proceedings of the 33rd International Conference on
  Software Engineering}} (Waikiki, Honolulu, HI, USA)
  \emph{(\bibinfo{series}{ICSE '11})}. \bibinfo{publisher}{ACM},
  \bibinfo{address}{New York, NY, USA}, \bibinfo{pages}{11--20}.
\newblock
\showISBNx{978-1-4503-0445-0}
\urldef\tempurl%
\url{https://doi.org/10.1145/1985793.1985796}
\showDOI{\tempurl}


\bibitem[\protect\citeauthoryear{{Tan}, {Marinov}, {Tan}, and {Leavens}}{{Tan}
  et~al\mbox{.}}{2012}]%
        {tcomment}
\bibfield{author}{\bibinfo{person}{S.~H. {Tan}}, \bibinfo{person}{D.
  {Marinov}}, \bibinfo{person}{L. {Tan}}, {and} \bibinfo{person}{G.~T.
  {Leavens}}.} \bibinfo{year}{2012}\natexlab{}.
\newblock \showarticletitle{@tComment: Testing Javadoc Comments to Detect
  Comment-Code Inconsistencies}. In \bibinfo{booktitle}{\emph{2012 IEEE Fifth
  International Conference on Software Testing, Verification and Validation}}.
  \bibinfo{pages}{260--269}.
\newblock
\showISSN{2159-4848}
\urldef\tempurl%
\url{https://doi.org/10.1109/ICST.2012.106}
\showDOI{\tempurl}


\bibitem[\protect\citeauthoryear{Tizpaz-Niari, {\v{C}}ern{\`y}, and
  Trivedi}{Tizpaz-Niari et~al\mbox{.}}{2020}]%
        {perf-bug-ML}
\bibfield{author}{\bibinfo{person}{Saeid Tizpaz-Niari}, \bibinfo{person}{Pavol
  {\v{C}}ern{\`y}}, {and} \bibinfo{person}{Ashutosh Trivedi}.}
  \bibinfo{year}{2020}\natexlab{}.
\newblock \showarticletitle{Detecting and understanding real-world differential
  performance bugs in machine learning libraries}. In
  \bibinfo{booktitle}{\emph{Proceedings of the 29th ACM SIGSOFT International
  Symposium on Software Testing and Analysis}}. \bibinfo{pages}{189--199}.
\newblock


\bibitem[\protect\citeauthoryear{Udeshi and Chattopadhyay}{Udeshi and
  Chattopadhyay}{2019}]%
        {grammar-based-testing}
\bibfield{author}{\bibinfo{person}{Sakshi Udeshi} {and}
  \bibinfo{person}{Sudipta Chattopadhyay}.} \bibinfo{year}{2019}\natexlab{}.
\newblock \showarticletitle{Grammar Based Directed Testing of Machine Learning
  Systems}.
\newblock \bibinfo{journal}{\emph{CoRR}}  \bibinfo{volume}{abs/1902.10027}
  (\bibinfo{year}{2019}).
\newblock
\showeprint[arxiv]{1902.10027}


\bibitem[\protect\citeauthoryear{Vanover, Deng, and Rubio-González}{Vanover
  et~al\mbox{.}}{2020}]%
        {discrepancies}
\bibfield{author}{\bibinfo{person}{Jackson Vanover}, \bibinfo{person}{Xuan
  Deng}, {and} \bibinfo{person}{Cindy Rubio-González}.}
  \bibinfo{year}{2020}\natexlab{}.
\newblock \showarticletitle{Discovering discrepancies in numerical libraries.}.
  In \bibinfo{booktitle}{\emph{Proceedings of the 2020 International Symposium
  on Software Testing and Analysis}} \emph{(\bibinfo{series}{ISSTA 2020})}.
  \bibinfo{pages}{488--501}.
\newblock
\urldef\tempurl%
\url{https://doi.org/10.1145/3395363.3397380}
\showDOI{\tempurl}


\bibitem[\protect\citeauthoryear{Wang, Sun, and Xing}{Wang
  et~al\mbox{.}}{2019}]%
        {wang2018simply}
\bibfield{author}{\bibinfo{person}{Haohan Wang}, \bibinfo{person}{Da Sun},
  {and} \bibinfo{person}{Eric~P Xing}.} \bibinfo{year}{2019}\natexlab{}.
\newblock \showarticletitle{What if we simply swap the two text fragments? a
  straightforward yet effective way to test the robustness of methods to
  confounding signals in nature language inference tasks}. In
  \bibinfo{booktitle}{\emph{Proceedings of the AAAI Conference on Artificial
  Intelligence}}, Vol.~\bibinfo{volume}{33}. \bibinfo{pages}{7136--7143}.
\newblock


\bibitem[\protect\citeauthoryear{Wang, Yan, Chen, Liu, and Zhang}{Wang
  et~al\mbox{.}}{2020}]%
        {wang:fse20}
\bibfield{author}{\bibinfo{person}{Zan Wang}, \bibinfo{person}{Ming Yan},
  \bibinfo{person}{Junjie Chen}, \bibinfo{person}{Shuang Liu}, {and}
  \bibinfo{person}{Dongdi Zhang}.} \bibinfo{year}{2020}\natexlab{}.
\newblock \showarticletitle{Deep Learning Library Testing via Effective Model
  Generation}. In \bibinfo{booktitle}{\emph{Proceedings of the 2020 28th ACM
  Joint Meeting on European Software Engineering Conference and Symposium on
  the Foundations of Software Engineering}} \emph{(\bibinfo{series}{ESEC/FSE
  2020})}.
\newblock


\bibitem[\protect\citeauthoryear{Wong, Zhang, Wang, Liu, and Tan}{Wong
  et~al\mbox{.}}{2015}]%
        {wong2015dase}
\bibfield{author}{\bibinfo{person}{Edmund Wong}, \bibinfo{person}{Lei Zhang},
  \bibinfo{person}{Song Wang}, \bibinfo{person}{Taiyue Liu}, {and}
  \bibinfo{person}{Lin Tan}.} \bibinfo{year}{2015}\natexlab{}.
\newblock \showarticletitle{Dase: Document-assisted symbolic execution for
  improving automated software testing}. In \bibinfo{booktitle}{\emph{2015
  IEEE/ACM 37th IEEE International Conference on Software Engineering}},
  Vol.~\bibinfo{volume}{1}. IEEE, \bibinfo{pages}{620--631}.
\newblock


\bibitem[\protect\citeauthoryear{Wu, Wu, Liang, Wang, Xie, and Mei}{Wu
  et~al\mbox{.}}{2013}]%
        {indicator}
\bibfield{author}{\bibinfo{person}{Qian Wu}, \bibinfo{person}{Ling Wu},
  \bibinfo{person}{Guangtai Liang}, \bibinfo{person}{Qianxiang Wang},
  \bibinfo{person}{Tao Xie}, {and} \bibinfo{person}{Hong Mei}.}
  \bibinfo{year}{2013}\natexlab{}.
\newblock \showarticletitle{Inferring dependency constraints on parameters for
  web services}. In \bibinfo{booktitle}{\emph{Proceedings of the 22nd
  international conference on World Wide Web}}. \bibinfo{pages}{1421--1432}.
\newblock


\bibitem[\protect\citeauthoryear{Xie, Ma, Juefei-Xu, Xue, Chen, Liu, Zhao, Li,
  Yin, and See}{Xie et~al\mbox{.}}{2019}]%
        {deep-hunter}
\bibfield{author}{\bibinfo{person}{Xiaofei Xie}, \bibinfo{person}{Lei Ma},
  \bibinfo{person}{Felix Juefei-Xu}, \bibinfo{person}{Minhui Xue},
  \bibinfo{person}{Hongxu Chen}, \bibinfo{person}{Yang Liu},
  \bibinfo{person}{Jianjun Zhao}, \bibinfo{person}{Bo Li},
  \bibinfo{person}{Jianxiong Yin}, {and} \bibinfo{person}{Simon See}.}
  \bibinfo{year}{2019}\natexlab{}.
\newblock \showarticletitle{DeepHunter: A Coverage-guided Fuzz Testing
  Framework for Deep Neural Networks}. In \bibinfo{booktitle}{\emph{Proceedings
  of the 28th ACM SIGSOFT International Symposium on Software Testing and
  Analysis}} (Beijing, China) \emph{(\bibinfo{series}{ISSTA 2019})}.
  \bibinfo{publisher}{ACM}, \bibinfo{address}{New York, NY, USA},
  \bibinfo{pages}{146--157}.
\newblock
\showISBNx{978-1-4503-6224-5}
\urldef\tempurl%
\url{https://doi.org/10.1145/3293882.3330579}
\showDOI{\tempurl}


\bibitem[\protect\citeauthoryear{Zaki}{Zaki}{2005}]%
        {subtreemining}
\bibfield{author}{\bibinfo{person}{Mohammed~Javeed Zaki}.}
  \bibinfo{year}{2005}\natexlab{}.
\newblock \showarticletitle{Efficiently mining frequent trees in a forest:
  Algorithms and applications}.
\newblock \bibinfo{journal}{\emph{IEEE transactions on knowledge and data
  engineering}} \bibinfo{volume}{17}, \bibinfo{number}{8}
  (\bibinfo{year}{2005}), \bibinfo{pages}{1021--1035}.
\newblock


\bibitem[\protect\citeauthoryear{Zhai, Shi, Pan, Zhou, Liu, Fang, Ma, Tan, and
  Zhang}{Zhai et~al\mbox{.}}{2020}]%
        {c2s}
\bibfield{author}{\bibinfo{person}{Juan Zhai}, \bibinfo{person}{Yu Shi},
  \bibinfo{person}{Minxue Pan}, \bibinfo{person}{Guian Zhou},
  \bibinfo{person}{Yongxiang Liu}, \bibinfo{person}{Chunrong Fang},
  \bibinfo{person}{Shiqing Ma}, \bibinfo{person}{Lin Tan}, {and}
  \bibinfo{person}{Xiangyu Zhang}.} \bibinfo{year}{2020}\natexlab{}.
\newblock \showarticletitle{C2S: Translating Natural Language Comments to
  Formal Program}. In \bibinfo{booktitle}{\emph{Proceedings of the 2020 28th
  ACM Joint Meeting on European Software Engineering Conference and Symposium
  on the Foundations of Software Engineering}} \emph{(\bibinfo{series}{ESEC/FSE
  2020})}.
\newblock


\bibitem[\protect\citeauthoryear{Zhang, Xiao, Zhang, Liu, Lin, and Yang}{Zhang
  et~al\mbox{.}}{2020}]%
        {zhang2020an}
\bibfield{author}{\bibinfo{person}{Ru Zhang}, \bibinfo{person}{Wencong Xiao},
  \bibinfo{person}{Hongyu Zhang}, \bibinfo{person}{Yu Liu},
  \bibinfo{person}{Haoxiang Lin}, {and} \bibinfo{person}{Mao Yang}.}
  \bibinfo{year}{2020}\natexlab{}.
\newblock \showarticletitle{An Empirical Study on Program Failures of Deep
  Learning Jobs}. In \bibinfo{booktitle}{\emph{Proceedings of the 42nd
  International Conference on Software Engineering (ICSE '20)}}.
  \bibinfo{publisher}{IEEE/ACM}.
\newblock


\bibitem[\protect\citeauthoryear{Zhang, Chen, Cheung, Xiong, and Zhang}{Zhang
  et~al\mbox{.}}{2018}]%
        {tf_bugs}
\bibfield{author}{\bibinfo{person}{Yuhao Zhang}, \bibinfo{person}{Yifan Chen},
  \bibinfo{person}{Shing-Chi Cheung}, \bibinfo{person}{Yingfei Xiong}, {and}
  \bibinfo{person}{Lu Zhang}.} \bibinfo{year}{2018}\natexlab{}.
\newblock \showarticletitle{An empirical study on TensorFlow program bugs}. In
  \bibinfo{booktitle}{\emph{Proceedings of the 2020 International Symposium on
  Software Testing and Analysis}} \emph{(\bibinfo{series}{ISSTA 2018})}.
  \bibinfo{pages}{129--140}.
\newblock
\urldef\tempurl%
\url{https://doi.org/10.1145/3213846.3213866}
\showDOI{\tempurl}


\bibitem[\protect\citeauthoryear{{Zheng}, {Wang}, {Liu}, {Zhang}, {Zeng},
  {Deng}, {Yang}, {He}, and {Xie}}{{Zheng} et~al\mbox{.}}{2019}]%
        {8802818}
\bibfield{author}{\bibinfo{person}{W. {Zheng}}, \bibinfo{person}{W. {Wang}},
  \bibinfo{person}{D. {Liu}}, \bibinfo{person}{C. {Zhang}}, \bibinfo{person}{Q.
  {Zeng}}, \bibinfo{person}{Y. {Deng}}, \bibinfo{person}{W. {Yang}},
  \bibinfo{person}{P. {He}}, {and} \bibinfo{person}{T. {Xie}}.}
  \bibinfo{year}{2019}\natexlab{}.
\newblock \showarticletitle{Testing Untestable Neural Machine Translation: An
  Industrial Case}. In \bibinfo{booktitle}{\emph{2019 IEEE/ACM 41st
  International Conference on Software Engineering: Companion Proceedings
  (ICSE-Companion)}}. \bibinfo{pages}{314--315}.
\newblock


\bibitem[\protect\citeauthoryear{Zhong, Zhang, Xie, and Mei}{Zhong
  et~al\mbox{.}}{2009}]%
        {Doc2Spec}
\bibfield{author}{\bibinfo{person}{Hao Zhong}, \bibinfo{person}{Lu Zhang},
  \bibinfo{person}{Tao Xie}, {and} \bibinfo{person}{Hong Mei}.}
  \bibinfo{year}{2009}\natexlab{}.
\newblock \showarticletitle{Inferring resource specifications from natural
  language API documentation}. In \bibinfo{booktitle}{\emph{2009 IEEE/ACM
  International Conference on Automated Software Engineering}}. IEEE,
  \bibinfo{pages}{307--318}.
\newblock


\bibitem[\protect\citeauthoryear{Zhou, Wang, Yan, Chen, Panichella, and
  Gall}{Zhou et~al\mbox{.}}{2018}]%
        {zhou2017analyzing}
\bibfield{author}{\bibinfo{person}{Yu Zhou}, \bibinfo{person}{Changzhi Wang},
  \bibinfo{person}{Xin Yan}, \bibinfo{person}{Taolue Chen},
  \bibinfo{person}{Sebastiano Panichella}, {and} \bibinfo{person}{Harald
  Gall}.} \bibinfo{year}{2018}\natexlab{}.
\newblock \showarticletitle{Automatic detection and repair recommendation of
  directive defects in Java API documentation}.
\newblock \bibinfo{journal}{\emph{IEEE Transactions on Software Engineering}}
  \bibinfo{volume}{46}, \bibinfo{number}{9} (\bibinfo{year}{2018}),
  \bibinfo{pages}{1004--1023}.
\newblock


\end{thebibliography}

\end{document}